\documentclass[aps,prd,preprint,groupedaddress,preprintnumbers,longbibliography,nofootinbib]{revtex4-1}

\newcommand{\nn}{\nonumber}
\newcommand{\lb}{\left\lbrace}
\newcommand{\rb}{\right\rbrace}
\newcommand{\Op}{\mathcal{O}}
\newcommand{\op}[1]{\left[ #1 \right]}

\newcommand{\PP}{\mathcal{P}}
\newcommand{\JJ}{\mathcal{J}}
\newcommand{\vev}[1]{\left\langle #1 \right\rangle}
\newcommand{\vvev}[1]{\left\langle\kern-0.3em\left\langle #1
    \right\rangle\kern-0.3em\right\rangle}

\usepackage{amsmath}
\usepackage{graphicx}
\usepackage{amsfonts}

\begin{document}


\preprint{KOBE-TH-17-04}

\title{Products of composite operators \\in the exact renormalization
  group formalism}


\author{C.~Pagani}
\email[]{capagani@uni-mainz.de}
\affiliation{Institute f\"{u}r Physik (WA THEP)
  Johannes-Gutenberg-Universit\"{a}t\\ Staudingerweg 7, 55099 Mainz,
  Germany}
\author{H.~Sonoda}\email[]{hsonoda@kobe-u.ac.jp}\affiliation{Physics
  Department, Kobe University, Kobe 657-8501, Japan}


\date{\today}

\begin{abstract}
  We discuss a general method of constructing the products of
  composite operators using the exact renormalization group formalism.
 Considering mainly the Wilson action at a generic fixed point of the
  renormalization group, we give an argument for the validity of short
  distance expansions of operator products.  We show how to compute
  the expansion coefficients by solving differential equations, and
  test our method with some simple examples.
\end{abstract}


\maketitle

\section{Introduction}

In the framework of the Wilsonian renormalization group (RG), the
physics of a system is completely characterized by a Wilson action.
The momentum cutoff of the action is fixed by rescaling while the
corresponding size in physical units diminishes exponentially under
the RG transformation.  The Wilson action of a critical theory
eventually reaches a fixed point which is scale invariant with no
characteristic length.  It is important to understand how small
deformations of the fixed point Wilson action grow under the RG
transformation.  For example, the exponential growth of deformations
is dictated by the critical exponents.  There are also space dependent
deformations with nontrivial rotation properties.  These deformations
constitute what we call composite operators, and their scaling
properties under the RG transformation constitute an essential part of
our understanding of critical phenomena and continuum field theory.

The purpose of this paper is to improve our understanding of composite
operators using the formalism of the exact renormalization group (ERG)
or functional RG.  The importance of composite operators in ERG was
emphasized early by Becchi \cite{Becchi:1996an}, and his results have
been extended in some later works such as
\cite{Pawlowski:2005xe,Igarashi:2009tj,Pagani:2016pad}.

In this work we use ERG to construct products of composite operators
and study their properties.  We discuss the insertion of two (or more)
composite operators 
in correlation functions of the elementary fields.  Particular
attention is paid to the ERG differential equation satisfied by
composite operators and their products at the fixed point.

When considering the product of two composite operators, it is natural
to ask about its short distance behavior.  At short distances the
operator product expansion (OPE) of K.~Wilson \cite{Wilson:1969zs} is
expected to be valid.  In the past ERG has been used to provide an
alternative proof of the existence of the OPE in perturbation
theory~\cite{Hughes:1988cp,Keller:1991bz,Keller:1992by,Hollands:2011gf,Holland:2014ifa,Holland:2014pna,Frob:2015uqy,Frob:2016mzv};
the original perturbative proof goes back to
Zimmermann~\cite{Zimmermann:1972tv}.  The purpose of examining the OPE
within ERG is to fill the gap between ERG and other nonperturbative
approaches to quantum field theory where the OPE forms the backbone
structure of the theory.  Particularly relevant is the case of
conformal field theories, especially in the two dimensional case.
Using ERG we argue the plausibility (if not a proof) of the existence
of OPE.  In particular, we derive ERG differential equations (a.k.a. flow
equations) satisfied by the Wilson coefficients and solve them for
some simple examples.

The paper is organized as follows.  In section \ref{section-comp} we
define composite operators at a fixed point of the RG transformation.
We introduce three equivalent approaches using the Wilson action
\cite{Wilson:1973jj,Polchinski:1983gv}, the generating functional of
connected correlation functions with an infrared cutoff
\cite{Morris:1993qb}, and its Legendre transform (called the effective
average action) \cite{Wetterich:1992yh,Ellwanger:1993mw},
respectively.  The three approaches differ in the natural choice of
field variables: $\phi, J, \Phi$.  In section \ref{section-product} we
generalize our construction to the product of two composite operators
and consider how the OPE arises.  In section \ref{section-examples}
some working examples are presented.  In section
\ref{section-multiple} we explain how to generalize the ERG
differential equations to consider the insertion of an arbitrary
number of composite operators, and in section \ref{section-massive} we
discuss the ERG differential equations for composite operators away
from the fixed point.  We summarize our findings in section
\ref{section-conclusions}.  We confine some technical parts in three
appendices.  In Appendix \ref{appendix-ERG} we review the basics of
the ERG formalism that this paper is based on.  A best pedagogical
effort has been made for those readers familiar with
\cite{Polchinski:1983gv} but not with \cite{Wilson:1973jj}.  In
Appendix \ref{appendix-Gaussian} we explain how to construct local
composite operators in the massive free scalar theories.  In Appendix
\ref{appendix-F} we derive the asymptotic behavior of a short-range
function necessary for the examples of Sec.~\ref{section-examples}.

We shall work in the dimensionless convention, where all dimensionful
quantities have been rescaled via a suitable power of the cutoff.  We
also adopt the following notation;
\begin{equation}
  \int_p = \int \frac{d^D p}{(2 \pi)^D}\,,\quad
  \delta (p) = (2\pi)^D \delta^{(D)} (p)\,,\quad 
  p \cdot \partial_p = \sum_{\mu=1}^D p_\mu \frac{\partial}{\partial p_\mu}\,.
\end{equation}

\section{Composite operators at a fixed point\label{section-comp}}

At a fixed point of the exact renormalization group, the Wilson action
satisfies the ERG equation\cite{Wilson:1973jj, Igarashi:2016qdr}
\begin{eqnarray}
0 &=& \int_p \left[ \left(- p \cdot \partial_p \ln K(p) + \frac{D+2}{2} -
      \gamma + p \cdot \partial_p \right) \phi (p) \cdot
    \frac{\delta}{\delta \phi (p)} \right.\nn\\
&& \left.\quad + \left( - p \cdot \partial_p \ln R(p) + 2 - 2
   \gamma \right) \frac{K(p)^2}{R(p)} \frac{1}{2} \frac{\delta^2}{\delta
   \phi (p) \delta \phi (-p)} \right] \, e^{S [\phi]}\,,\label{ERG-S}
\end{eqnarray}
where $\gamma$ is the anomalous dimension.  (We have prepared Appendix
\ref{appendix-ERG} for the readers who are familiar with
\cite{Polchinski:1983gv} but not with \cite{Wilson:1973jj}.)  We have
introduced two positive cutoff functions:
\begin{enumerate}
\item $K(p)$ approaches $1$ as $p^2 \to 0$, and decays rapidly for
  $p^2 \gg 1$.
\item $R(p)$ must be nonvanishing at $p = 0$  and
  decays rapidly for $p^2 \gg 1$.  The inverse transform of $R(p)$ is
  a function in space that is nonvanishing only over a region of unit
  size.
\end{enumerate}
For example, the choice
$K(p) = e^{- p^2}, R(p) = \frac{p^2 K(p)}{1-K(p)} =
\frac{p^2}{e^{p^2}-1}$
satisfies the criteria.  The original choice made in
\cite{Wilson:1973jj} is $K(p) = e^{-p^2}, R(p) = e^{- 2 p^2}$.

Eq.~(\ref{ERG-S}) implies that the modified correlation functions
defined by
\begin{equation}
\vvev{\phi (p_1) \cdots \phi (p_n)}
\equiv \prod_{i=1}^n \frac{1}{K(p_i)} \cdot \vev{\exp \left( -
    \int_p \frac{K(p)^2}{R(p)} \frac{1}{2} \frac{\delta^2}{\delta \phi
      (p) \delta \phi (-p)} \right)\, \phi (p_1) \cdots \phi (p_n)}_S
\end{equation}
satisfy the scaling law
\begin{equation}
\vvev{\phi (p_1 e^t) \cdots \phi (p_n e^t)} = \exp \left( n \left(-
    \frac{D+2}{2} + \gamma \right)\right) \vvev{\phi (p_1) \cdots \phi
  (p_n)}
\end{equation}
for arbitrary momenta.\cite{Sonoda:2015bla}

In our discussion of composite operators we find it more convenient to
deal with a functional $W[J]$ defined directly in terms of $S[\phi]$
as\cite{Morris:1993qb, Morris:1994ie}
\begin{subequations}
\label{WJ-def}
\begin{eqnarray}
W [J] &\equiv& \frac{1}{2} \int_p \frac{J (p) J(-p)}{R(p)} + S
[\phi]\,,\label{W-def}\\
\textrm{where}\quad J (p) &\equiv& \frac{R (p)}{K(p)} \phi (p)\,.\label{J-def}
\end{eqnarray}
\end{subequations}
In fact it is even more convenient to deal with the Legendre transform
of $W[J]$\cite{Wetterich:1992yh, Berges:2000ew}:
\begin{subequations}
\label{GammaPhi-def}
\begin{eqnarray}
- \frac{1}{2} \int_p R (p) \Phi (p) \Phi (-p) + \Gamma [\Phi] &\equiv&
W[J] - \int_p J(-p) \Phi (p)\,,\label{Gamma-def}\\
\Phi (p) &\equiv& \frac{\delta W[J]}{\delta J(-p)}\,.\label{Phi-def}
\end{eqnarray}
\end{subequations}
$\Gamma [\Phi]$ is often called the effective average action.  We can
interpret $W [J]$ as the generating functional of connected
correlation functions\cite{Morris:1993qb,Morris:1994ie} and $\Gamma
[\Phi]$ as the effective action\cite{Wetterich:1992yh,Berges:2000ew},
both in the presence of an infrared cutoff.  (The same cutoff is
called an ultraviolet cutoff for $S$ and an infrared cutoff for $W$
and $\Gamma$.  This is because we regard $S$ as the weight of
functional integration over low momenta to be done, but we regard $W$
and $\Gamma$ as consequences of functional integration over high
momenta already done.  It has recently been shown that the high
momentum limit of $W$ and $\Gamma$ gives the corresponding functionals
without the infrared cutoff.\cite{Sonoda:2017rro}) The ERG equations
satisfied by $W$ and $\Gamma$ are given by (see
\cite{Igarashi:2016qdr} and reference therein)
\begin{eqnarray}
0&=&\int_p J(-p) \left(- p \cdot \partial_p - \frac{D+2}{2} + \gamma
\right) \frac{\delta}{\delta J(-p)}\, e^{W[J]}\nn\\
&&\quad + \int_p \left( - p \cdot \partial_p + 2 - 2 \gamma \right)
R(p) \cdot \frac{1}{2} \frac{\delta^2}{\delta J(p) \delta J(-p)} e^{W
  [J]}\,,\\
0&=& \int_p \frac{\delta \Gamma [\Phi]}{\delta \Phi (p)} \left(- p
  \cdot \partial_p - \frac{D+2}{2} + \gamma
\right) \Phi (p)\nn\\
&& \quad + \int_p \left( - p \cdot \partial_p + 2 - 2 \gamma \right) R(p)
\cdot \frac{1}{2} G_{p, -p} [\Phi]\,, 
\end{eqnarray}
where
\begin{equation}
G_{p, q} [\Phi] \equiv \frac{\delta^2 W[J]}{\delta J(p) \delta J(q)}
\label{G-def}
\end{equation}
satisfies
\begin{equation}
\int_q G_{p,q} [\Phi] \left( R (q) \delta (q-r) - \frac{\delta^2
    \Gamma [\Phi]}{\delta \Phi (-q) \delta \Phi (r)} \right) = \delta
(p-r)\,.
\end{equation}

Now, composite operators can be thought of as infinitesimal changes of
$S$, $W$, or $\Gamma$.  Correspondingly, we can regard composite
operators as functionals of $\phi$, $J$, or $\Phi$.  Let $\Op (p)$ be
a composite operator of scale dimension $-y$ and momentum $p$.
Regarding it as a functional of $J$, we obtain the following ERG
equation:
\begin{eqnarray}
&&\left( y + p \cdot \partial_p \right) \Op (p)
= \int_q \lb J(q) \left(- q \cdot \partial_q - \frac{D+2}{2} + \gamma
\right) \frac{\delta}{\delta J(q)}\right.\nn\\
&&\quad \left.\quad + \left(- q \cdot \partial_q + 2 - 2\gamma \right) R(q)
\left(
\frac{\delta W[J]}{\delta J(-q)} \frac{\delta}{\delta J(q)} +
\frac{1}{2} \frac{\delta^2}{\delta J(q) \delta J(-q)}\right) \rb \,
\Op (p)\,.
\label{ERG-OpW}
\end{eqnarray}
Similarly, regarding $\Op (p)$ as a functional of $\Phi$, we can
rewrite the above as
\begin{eqnarray}
&&\left( y + p \cdot \partial_p \right) \Op (p)
= \int_q \lb \left(  q \cdot \partial_q + \frac{D+2}{2} - \gamma
\right) \Phi (q) \cdot \frac{\delta}{\delta \Phi (q)}\right.\nn\\
&&\left.\quad\quad + \left(- q \cdot \partial_q + 2 - 2\gamma \right) R(q)
\frac{1}{2} \int_{r,s} G_{q,-r} [\Phi] G_{-q,-s} [\Phi]
\frac{\delta^2}{\delta \Phi (r) \delta \Phi (s)} \rb \Op (p)\,,
\label{ERG-OpGamma}
\end{eqnarray}
where $G$ is defined by (\ref{G-def}). 

The above two ERG equations are equivalent, and they imply that the
modified correlation functions defined by
\begin{eqnarray}
&&\vvev{\Op (p)\, \phi (p_1) \cdots \phi (p_n)}\nn\\
&&\equiv \prod_{i=1}^n \frac{1}{K(p_i)}\, \vev{\Op (p) \exp\left( -
    \int_q \frac{K(q)^2}{R(q)} \frac{1}{2} \frac{\delta^2}{\delta \phi
      (q) \delta \phi (-q)} \right)\, \phi (p_1) \cdots \phi (p_n)}_S
\end{eqnarray}
satisfy the scaling law
\begin{equation}
\vvev{\Op (p e^t)\, \phi (p_1 e^t) \cdots \phi (p_n e^t)}
= \exp \left( t \left( - y - n \frac{D+2}{2} \right) \right) \vvev{\Op
  (p)\, \phi (p_1) \cdots \phi (p_n)}\,.
\end{equation}

\section{Products of composite operators\label{section-product}}

Given two composite operators $\Op_1 (p), \Op_2 (p)$ of scale
dimensions $- y_1, - y_2$, we wish to define their product as a
composite operator of scale dimension $- (y_1+y_2)$.  The naive
product $\Op_1 (p) \Op_2 (q)$ will not do because it does not satisfy
(\ref{ERG-OpW}) or (\ref{ERG-OpGamma}).  We must define the product by
adding a \textbf{local} counterterm:
\begin{equation}
\left[ \Op_1 (p) \Op_2 (q) \right] \equiv \Op_1 (p) \Op_2 (q) +
\PP_{12} (p,q)\,.\label{product}
\end{equation}
Otherwise the product will not satisfy the scaling law:
\begin{eqnarray}
&&\vvev{\left[\Op_1 (p e^t) \Op_2 (q e^t)\right] \phi (p_1 e^t) \cdots
  \phi (p_n e^t)}\nn\\
&&= \exp \left[ t \left( - y_1 - y_2 + n \left( - \frac{D+2}{2} +
      \gamma \right) \right) \right]
\vvev{\left[ \Op_1 (p) \Op_2 (q) \right] \phi (p_1) \cdots \phi
  (p_n)}\,.\label{comp-scaling}
\end{eqnarray}

In the ERG formalism we have a dimensionless cutoff of order $1$
(either in momentum space or in coordinate space).  In coordinate
space the inverse Fourier transform
\begin{equation}
\Op (r) = \int_p e^{i p r} \Op (p)
\end{equation}
is expected to have a support of unit size around the coordinate
$r$.  We expect the same property for the product of two composite
operators.  Given $\Op_1 (p)$ and $\Op_2 (q)$, we denote their inverse
Fourier transforms using the same symbol:
\begin{equation}
\lb\begin{array}{c@{~=~}l}
\Op_1 (r) & \int_p e^{i p r} \Op_1 (p)\,,\\
\Op_2 (r') & \int_q  e^{ i q r'} \Op_2 (q)\,.
\end{array}\right.
\end{equation}
Both have a distribution of unit size in coordinate space.  The limit
\begin{equation}
\Op_1 (r) \Op_2 (r') \overset{r' \to r}{\longrightarrow} \Op_1 (r) \Op_2 (r)
\end{equation}
is well defined. If there are short-distance singularities, we cannot
find them in $\Op_1 (p) \Op_2 (q)$: we must look for them in the
counterterm $\PP_{12} (p,q)$, which is required by ERG (or
equivalently scaling).  Even without the help of ERG, we expect the
need for the counterterm in defining the Fourier transform; the
integration over the case where the two operators are dangerously
close together requires special attention, resulting in a local
counterterm.

Let us further analyze the nature of $\PP_{12} (p,q)$.  Regarding
composite operators as functionals of $\Phi$, we obtain
\begin{subequations}
\label{ERG-O1O2}
\begin{eqnarray}
\left(y_1 + p \cdot \partial_p - \mathcal{D} \right) \Op_1 (p) &=&
0\,,\\
\left(y_2 + q \cdot \partial_q - \mathcal{D} \right) \Op_2 (q) &=&
0\,,\\
\left(y_1 + y_2 + p \cdot \partial_p + q \cdot \partial_q -
  \mathcal{D} \right) \left[ \Op_1 (p) \Op_2 (q)\right] &=& 0\,,
\label{ERG-product}
\end{eqnarray}
\end{subequations}
where
\begin{eqnarray}
\mathcal{D} &\equiv& \int_r \lb \left( r \cdot \partial_r +
  \frac{D+2}{2} - \gamma \right) \Phi (r) \cdot \frac{\delta}{\delta
  \Phi (r)}\right.\nn\\
&&\left.\quad + \left(- r \cdot \partial_r +2 -2 \gamma \right) R(r)
  \, \frac{1}{2} \int_{s,t} G_{r,-s} [\Phi] G_{-r,-t} [\Phi]
  \frac{\delta^2}{\delta \Phi (s) \delta \Phi (t)} \rb\,.
\label{calD-def}
\end{eqnarray}
(\ref{ERG-product}) is equivalent to the scaling (\ref{comp-scaling}).

From (\ref{ERG-O1O2}), we obtain the following ERG equation for the
counterterm: 
\begin{eqnarray}
&& \left( y_1 + y_2 + p \cdot \partial_p + q \cdot \partial_q -
  \mathcal{D} \right) \PP_{12} (p,q)\nn\\
&&= \int_r \left(- r \cdot \partial_r +2 -2 \gamma \right) R(r)\cdot
  \, \int_s G_{r,-s} [\Phi] \frac{\delta \Op_1
    (p)}{\delta \Phi (s)} \int_t G_{-r,-t} [\Phi] \frac{\delta \Op_2
    (q)}{\delta \Phi (t)}\nn\\
&&= \int_r \left(- r \cdot \partial_r +2 -2 \gamma \right) R(r)\cdot
\frac{\delta \Op_1 (p)}{\delta J(r)} \frac{\delta \Op_2 (q)}{\delta
  J(-r)}\,, \label{ERG-PP}
\end{eqnarray}
where we have used
\begin{equation}
G_{r,-s} [\Phi] = \frac{\delta W[J]}{\delta J(r) \delta J(-s)} =
\frac{\delta \Phi (s)}{\delta J(r)}\,.
\end{equation}
Since $R$ is the Fourier transform of a function nonvanishing only
over a region of unit size, (\ref{ERG-PP}) is local in space.  That
means that the inverse Fourier transform
\[
\int_{p,q} e^{i p x + i q x'} \int_r \left(- r \cdot \partial_r + 2 -
  2 \gamma \right) R(r) \cdot \frac{\delta \Op_1 (p)}{\delta J(r)}
\frac{\delta \Op_2 (q)}{\delta J(-r)}
\]
is nonvanishing only when the distance $|x - x'|$ is of order $1$ or
less.

Therefore, we can expand the counterterm $\PP_{12} (p,q)$ using a
basis of local composite operators:
\begin{equation}
\PP_{12} (p,q) = \sum_i c_{12,i} (p-q) \Op_i (p+q)\,,\label{expand-PP}
\end{equation}
where $\Op_i$ is a composite operator of scale dimension $- y_i$,
satisfying
\begin{equation}
\left( y_i + p \cdot \partial_p - \mathcal{D} \right) \Op_i (p) = 0\,.
\end{equation}
The coefficient $c_{12,i}$ depends only on $p-q$; we have absorbed all
the dependence on $p+q$ into $\Op_i$.  Similarly, we can expand the
right-hand side of (\ref{ERG-PP}) as
\begin{equation}
\int_r \left( - r \cdot \partial_r + 2 - 2 \gamma \right) R(r) \cdot
\frac{\delta \Op_1 (p)}{\delta J (r)} 
\frac{\delta \Op_2 (q)}{\delta J (-r)} = \sum_i d_{12, i} (p-q)
\Op_i (p+q)\,.\label{ERG-PP-RHS}
\end{equation}
Substituting (\ref{expand-PP}) and (\ref{ERG-PP-RHS}) into
(\ref{ERG-PP}), we obtain
\[
 \left( p \cdot \partial_p + q \cdot \partial_q + y_1 + y_2 -
  y_i \right) c_{12,i} (p-q) = d_{12,i} (p-q)\,,
\]
or equivalently
\begin{equation}
  \left( p \cdot \partial_p + y_1 + y_2 -
    y_i \right) c_{12,i} (p) = d_{12,i} (p)
\label{diffeq-c}
\end{equation}
which determines $c_{12,i} (p)$ in terms of $d_{12, i} (p)$.

Before discussing the short-distance behavior of $c_{12,i} (p-q)$ for
large $|p-q|$, we would like to consider the solvability of
(\ref{diffeq-c}) and uniqueness of its solution.  We assume
analyticity: both $c_{12,i} (p-q)$ and $d_{12,i} (p)$ are regular
functions of $p$ at $p=0$.  (\ref{diffeq-c}) can be solved uniquely
unless
\begin{equation}
  n \equiv -(y_1 + y_2) + y_i = 0, 1, 2, \cdots\,.\label{degeneracy}
\end{equation}
If (\ref{degeneracy}) holds, and if $d_{12, i} (p)$ contains a
constant multiple of $p^n$, we cannot find an analytic solution.  (If
$c_{12,i} (p)$ is a scalar, the following discussion applies only for
even $n$.)  Even if $d_{12,i} (p)$ has no such a term, $c_{12,i} (p)$
is ambiguous by a constant multiple of $p^n$.  So, what to do if
(\ref{degeneracy}) is the case?

If (\ref{degeneracy}) holds, we need to modify (\ref{ERG-product}) so
that
\begin{equation}
\left( y_1 + y_2 + p \cdot \partial_p + q \cdot \partial_q -
  \mathcal{D} \right) \left[ \Op_1 (p) \Op_2 (q)\right] = d_i \cdot (p-q)^n
\Op_i (p+q)
\end{equation}
where the constant $d$ is determined so that
\[
d_{12,i} (p) + d_i \cdot p^n
\]
has no term proportional to $p^n$.  Then, we need to solve
\begin{equation}
  \left( p \cdot \partial_p - n \right) c_{12,i} (p) = d_{12,i} (p) +
  d_i \cdot p^n 
\label{diffeq-c-degeneracy}
\end{equation}
instead of (\ref{diffeq-c}).  This can be solved, but the solution is
not unique.  To fix the coefficient of $p^n$, we must introduce a
convention such as the absence of the $p^n$ term in $c_{12,i} (p)$:
\begin{equation}
\frac{\partial^n}{\partial p^n} c_{12,i} (p) \Big|_{p=0} = 0\,.
\end{equation}
All this implies that the scale dimension $y_1+y_2$ is extended to a
matrix: the product $\left[ \Op_1 (p) \Op_2 (q) \right]$ mixes with a
local operator $\Op_i (p+q)$ for which (\ref{degeneracy}) is
satisfied.  For example, at the Gaussian fixed point in $D=4$, the
product
\[
\left[\frac{1}{2} \left[ \phi^2 (p)\right]
\frac{1}{2} \left[ \phi^2 (q)\right]\right]
\]
($y_1=y_2=2$) mixes with $\delta (p+q)$.  See Example 1 of
Sec.~\ref{section-examples} for more details. So much for the
discussion of (\ref{degeneracy}).

Now, we consider the short-distance limit of the product.  We consider
$\left[\Op_1 (p) \Op_2 (q)\right]$, taking $|p-q|$ large while fixing
$p+q$.  As has been explained, a singular behavior is expected not of
$\Op_1 (p) \Op_2 (q)$ but of $\PP_{12} (p,q)$:
\begin{equation}
\vvev{\left[ \Op_1 (p) \Op_2 (q)\right] \phi (p_1) \cdots \phi (p_n)}
\overset{|p-q|\,\mathrm{large}}{\underset{p+q\,\mathrm{fixed}}{\approx}}
\vvev{\PP_{12} (p,q)\, \phi (p_1) \cdots \phi (p_n)}\,. 
\end{equation}
We can regard $\PP_{12} (p,q)$ as a functional of $J$ with momentum
$p+q$.  Since we keep the momenta $p_1, \cdots, p_n$ finite in the
above, we can assume $\frac{\delta \Op_1 (p)}{\delta J(r)}$ and
$\frac{\delta \Op_2 (q)}{\delta J(-r)}$ of (\ref{ERG-PP}) to depend
only on $J$'s with finite momenta.  Hence, $r$ in (\ref{ERG-PP}) must
be of order $p$ by momentum conservation.  Therefore, $R (r)$ becomes
extremely small.  Hence, from (\ref{ERG-PP-RHS}), we expect
\begin{equation}
  d_{12,i} (p-q) \overset{|p-q|\to
    \infty}{\longrightarrow} 0\,. 
\end{equation}
Thus, we obtain
\begin{equation}
 \left( p \cdot \partial_p + y_1 + y_2 -
  y_i \right) c_{12,i} (p-q) \overset{|p-q| \to
  \infty}{\longrightarrow} 0\,.
\end{equation}
This implies
\begin{equation}
c_{12,i} (p-q) \overset{|p-q| \to
  \infty}{\underset{p+q\,\mathrm{fixed}}{\longrightarrow}} 
 C_{12,i}\cdot p^{-y_1-y_2+y_i}\,,
\end{equation}
where $C_{12,i}$ is a constant.

We thus obtain a short-distance expansion (a.k.a. operator product
expansion) 
\begin{equation}
  \left[ \Op_1 (p) \Op_2 (q)\right]
  \overset{|p- q| \to
    \infty}{\underset{p+q\,\mathrm{fixed}}{\longrightarrow}}  \sum_i
  C_{12, i}\cdot   p^{-y_1-y_2+y_i} \,\Op_i (p+q)\,.
\end{equation}
To be able to neglect the contribution of $\Op_1 (p) \Op_2 (q)$, we
must restrict the sum over $i$ to
\begin{equation}
-y_1-y_2+y_i \ge - D\,,
\end{equation}
corresponding to singularities in space.  This condition can be
rewritten as
\begin{equation}
(D-y_1) + (D-y_2) \ge (D-y_i)\,,
\end{equation}
where $D-y_i$ is the scale dimension of the inverse Fourier transform
of $\Op_i$ (operator in coordinate space).  The operator $\Op_i$ with
the lowest scale dimension provides the highest short-distance singularity.

\section{Examples\label{section-examples}}

We would like to provide concrete applications of the general theory
we have developed.  In the first subsection we consider a generic
fixed point action, and in the second we take examples from the
Gaussian fixed point in $D$ dimensions ($2 < D \le 4$).

\subsection{$\left[\Op (p) \Phi (q)\right]$}

$\Phi (p)$ is a composite operator corresponding to the elementary
field $\phi (p)$.  $\Phi (p)$ satisfies (\ref{ERG-OpGamma}) with scale
dimension
\begin{equation}
- y_\Phi = - \frac{D+2}{2} + \gamma\,.
\end{equation}
Let $\Op (p)$ be an arbitrary composite operator of scale dimension
$-y$.  Its product with $\Phi (q)$ must satisfy
\begin{equation}
\vvev{\left[ \Op (p) \Phi (q) \right] \phi (p_1) \cdots \phi (p_n)}
= \vvev{\Op (p) \phi (q) \phi (p_1) \cdots \phi (p_n)}\,.
\end{equation}
This gives\cite{Igarashi:2009tj}
\begin{eqnarray}
\left[ \Op (p) \Phi (q)\right] &=& \Op (p) \Phi (q) + 
\frac{K(q)}{R(q)} \frac{\delta \Op (p)}{\delta \phi (-q)}\nn\\
&=& \Op (p) \Phi (q) + \frac{\delta \Op (p)}{\delta J (-q)}\,.
\label{product-OPhi}
\end{eqnarray}
This implies the counterterm
\begin{equation}
\PP_{\Op \Phi} (p,q)  =  \frac{\delta \Op (p)}{\delta J(-q)} \,.\label{PP-OPhi}
\end{equation}

For the simplest case of $\Op = \Phi$, we use (\ref{Phi-def}) and
(\ref{PP-OPhi}) to obtain
\begin{eqnarray}
\op{\Phi (p) \Phi (q)} &=& \frac{\delta W[J]}{\delta J(-p)} \frac{\delta
  W[J]}{\delta J(-q)} + \frac{\delta^2 W[J]}{\delta J(-p) \delta
  J(-q)}\nn\\
&=& e^{-W[J]} \frac{\delta^2}{\delta J(-p) \delta J(-q)}
e^{W[J]}\,.\label{Phi-Phi} 
\end{eqnarray}
This generalizes to\cite{Rosten:2010vm}
\begin{equation}
\op{\Phi (p_1) \cdots \Phi (p_n)} = e^{- W[J]} \frac{\delta^n}{\delta
  J(-p_1) \cdots \delta J(-p_n)} e^{W [J]}\,,\label{multi-Phi}
\end{equation}
which can be checked to satisfy (\ref{product-OPhi}):
\begin{eqnarray}
&&e^{-W[J]} \frac{\delta^n}{\delta  J(-p_1) \cdots \delta J(-p_n)} e^{W
  [J]} = \frac{\delta
W[J]}{\delta J(-p_n)} \cdot  e^{-W[J]}\frac{\delta^{n-1}}{\delta
  J(-p_1) \cdots \delta J(-p_{n-1})} e^{W [J]}\nn\\
&&\qquad\qquad + \frac{\delta}{\delta J(-p_n)} \left(
  e^{-W[J]}\frac{\delta^{n-1}}{\delta 
  J(-p_1) \cdots \delta J(-p_{n-1})} e^{W [J]}\right) \,.
\end{eqnarray}

Using (\ref{G-def}) we can rewrite (\ref{Phi-Phi}) as
\begin{equation}
\left[ \Phi (p) \Phi (q) \right] = \Phi (p) \Phi (q) + G_{-q,-p}
[\Phi]\,.
\end{equation}
Hence, for $p+q$ fixed, we obtain
\begin{equation}
\left[ \Phi (p) \Phi (q) \right] \overset{|p-q| \to
  \infty}{\underset{p+q\,\mathrm{fixed}}{\approx}} G_{-p,-q} [\Phi]\,.
\end{equation}
From the scaling law
\begin{equation}
\vvev{\phi (p) \phi (q)} = \textrm{const}\, \frac{1}{p^{2
    (1-\gamma)}}\, \delta (p+q)\,,
\end{equation}
we obtain the coefficient of the identity operator as
\begin{equation}
\left[ \Phi (p) \Phi (q) \right] 
\overset{|p-q| \to \infty}{\underset{p+q\,\mathrm{fixed}}{\longrightarrow}}
\textrm{const}\, \frac{1}{p^{2 (1-\gamma)}} \,\delta (p+q)\,.
\end{equation}
Further coefficients can be computed by employing some approximation
scheme.  For instance, we have checked in the $\phi^4$ theory in
dimension $D=4$ that the order $\lambda$ correction 
is given by
\begin{equation}
\op{\Phi (p) \Phi (q)} \overset{|p-q| \to
  \infty}{\underset{p+q\,\mathrm{fixed}}{\longrightarrow}}
\frac{1}{p^2} \delta (p+q) - \lambda \frac{1}{p^4} \frac{1}{2}
\op{\phi^2 (p+q)}\,.
\end{equation}
In coordinate space the order $\lambda$ correction is proportional to
the logarithm of the distance.

\subsection{Examples from the Gaussian fixed point in $D$ dimensions}

We now consider the composite operators at the Gaussian fixed point:
\begin{equation}
\Gamma [\Phi] = - \frac{1}{2} \int_p p^2 \Phi (p) \Phi (-p)\,.
\end{equation}
There is no anomalous dimension: $\gamma = 0$.  The high-momentum
propagator is given by
\begin{equation}
G_{p,q} [\Phi]  =  \frac{1}{p^2 + R(p)} \, \delta (p+q) \equiv h (p)
\delta (p+q)\,.
\end{equation}
For convenience we introduce
\begin{equation}
f(p) \equiv \left(2 + p \cdot \partial_p \right) h(p) = \frac{(2 - p
  \cdot \partial_p) R(p)}{\left(p^2 + R (p)\right)^2}\,,
\end{equation}
which vanishes rapidly for $p \gg 1$.  Now, the ``differential''
operator $\mathcal{D}$ defined by (\ref{calD-def}) can be written as
\begin{equation}
\mathcal{D} \equiv \int_r \lb
\left(r \cdot \partial_r + \frac{D+2}{2}\right) \Phi (r) \cdot
\frac{\delta}{\delta \Phi (r)} + f(r) \frac{1}{2}
\frac{\delta^2}{\delta \Phi (r) \delta \Phi (-r)} \rb\,.
\end{equation}
In the remaining part of this section we consider the products of the
composite operators
\begin{eqnarray}
\frac{1}{2} \op{\phi^2 (p)} &\equiv& \frac{1}{2} \int_{p_1, p_2} \Phi
(p_1) \Phi (p_2)\, \delta (p_1+p_2-p) + \kappa_2 \cdot \delta (p)\,,\\
\frac{1}{4!} \op{\phi^4 (p)} &\equiv& \frac{1}{4!} \int_{p_1, \cdots,
  p_4} \Phi (p_1) \cdots \Phi (p_4)\, \delta (p_1 + \cdots + p_4 -
p)\nn\\
&& + \kappa_2 \, \frac{1}{2} \int_{p_1, p_2} \Phi
(p_1) \Phi (p_2)\, \delta (p_1+p_2-p) + \frac{1}{2} \kappa_2^2\,
\delta (p)\,,
\end{eqnarray}
where the constant $\kappa_2$ is defined by
\begin{equation}
\kappa_2 \equiv - \frac{1}{D-2} \frac{1}{2} \int_q f(q)\,.
\end{equation}
The scale dimensions are $- 2$ and $D-4$, respectively.  See Appendix
\ref{appendix-Gaussian} for the construction of composite operators in
the free theory.

\subsection*{Example 1: $\left[\left[\frac{1}{2} \phi^2 (p)\right]
    \left[\frac{1}{2} \phi^2 (q)\right]\right]$}

The scale dimension of the product is $-y = - 4$.  Hence, in $D=4$, we
expect mixing with the unit operator $\delta (p+q)$.  Let
\begin{equation}
\left[ \frac{1}{2} \left[\phi^2 (p)\right] \frac{1}{2} \left[ \phi^2
    (q)\right]\right] =  \frac{1}{2}\left[\phi^2 (p)\right]
\frac{1}{2}\left[ \phi^2 (q)\right] + \PP_{22} (p,q)\,.
\end{equation}
The counterterm must satisfy
\begin{eqnarray}
&&\left(4 + p \cdot \partial_p + q \cdot \partial_q - \mathcal{D}\right)
\PP_{22} (p,q) \nn\\
&&= \int_r f(r) \frac{\delta}{\delta \Phi (r)} \frac{1}{2}
\left[\phi^2 (p)\right]  \frac{\delta}{\delta \Phi (-r)} \frac{1}{2}
\left[ \phi^2 (q)\right] \nn\\
&&= \int_r f(r) \Phi (p-r) \Phi (q+r)\,.\label{diffeq-P22}
\end{eqnarray}
To solve this, let us expand
\begin{eqnarray}
\PP_{22} (p,q) &=& \frac{1}{2} \int_{p_1, p_2} \Phi (p_1) \Phi (p_2)
\delta (p_1+p_2-p-q)\, u_2 (p-q; p_1, p_2) \nn\\
&& + u_0 (p) \delta (p+q)\,.
\end{eqnarray}
Substituting this into (\ref{diffeq-P22}), we obtain
\begin{subequations}
\begin{equation}
\left(2 + p \cdot \partial_p + q \cdot \partial_q + \sum_{i=1,2} p_i
  \cdot \partial_{p_i} \right) u_2 (p-q; p_1, p_2) =
f(p_1-p)+f(p_1-q)\,,
\label{diffeq-u2}
\end{equation}
and 
\begin{equation}
\left(4 - D + p\cdot \partial_p \right) u_0 (p) = \frac{1}{2} \int_r
f(r) u_2 (2 p; r,-r)\,.\label{diffeq-u0}
\end{equation}
\end{subequations}
The homogeneous solution of (\ref{diffeq-u2}) is excluded on account
of analyticity at zero momentum.  Hence, we obtain
\begin{equation}
u_2 (p-q; p_1, p_2) = h (p_1-p) + h (p_1 - q)\,.
\end{equation}
Substituting this into
(\ref{diffeq-u0}), we obtain
\begin{equation}
\left(4 - D + p\cdot \partial_p \right) u_0 (p) = \int_q f(q) h
(q+p)\,.
\label{diffeq-u0-2}
\end{equation}
For $2 < D < 4$, this is uniquely solved by
\begin{equation}
u_0 (p) = F (p) \equiv \frac{1}{2} \int_q h(q) h(q+p)\,.
\end{equation}
\begin{figure}[t]
\label{Fig1}
\centering
\includegraphics[width=8cm]{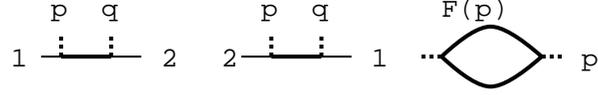}
\caption{Graphical representation of $\PP_{22} (p,q)$: a dark line
  for $h$}
\end{figure}
(See Fig. 1.)  Now, for $D=4$, (\ref{diffeq-u0-2}) does not admit a
solution analytic at $p=0$.  Since the left-hand side vanishes at
$p=0$, we must modify it to
\begin{equation}
p \cdot \partial_p u_0 (p) = \int_q f(q) \left( h (q+p) - h (q)
\right)\qquad (D=4)\label{diffeq-u0-D4}
\end{equation}
by subtracting a constant from the right-hand side.  The constant can
be evaluated as
\begin{eqnarray}
\int_q f(q) h(q) &=& \int_q h(q) (2+q \cdot \partial_q) h(q)\nn\\
&=&  \int_q (4 + q \cdot \partial_q) \frac{1}{2} h(q)^2\nn\\
&=& \frac{1}{2} \int_r \frac{\partial}{\partial q_\mu} \left( q_\mu
  h(q)^2 \right)\nn\\
&=& \frac{1}{(4 \pi)^2}\,.\label{integral-fh}
\end{eqnarray}
(\ref{diffeq-u0-D4}) determines $u_0 (p)$ up to an additive
constant.  We define $F(p)$ by
\begin{subequations}
\begin{eqnarray}
p \cdot \partial_p F(p) &=& \int_q f(q) h(q+p) -
\frac{1}{(4\pi)^2}\,,\label{F-4D-def}\\
F(0) &\equiv& 0\,.
\end{eqnarray}
\end{subequations}
As a convention we adopt the choice $u_0 (p) = F(p)$.  The subtraction
in (\ref{F-4D-def}) implies the mixing of the product with the
identity operator, and the product satisfies the ERG equation
\begin{equation}
\left( 4 + p \cdot \partial_p + q \cdot \partial_q - \mathcal{D}
\right) \op{ \frac{1}{2} \op{\phi^2 (p)} \frac{1}{2} \op{\phi^2 (q)} }
= - \frac{1}{(4\pi)^2} \delta (p+q)\,.
\end{equation}

Let us find the asymptotic behavior of the product as $p \to \infty$
for a fixed $p+q$.  We find
\begin{equation}
u_2 (p-q; p_1, p_2) \overset{|p-q| \to
  \infty}{\underset{p-q\,\mathrm{fixed}}{\longrightarrow}}
\frac{2}{p^2}\,.
\end{equation}
From Appendix \ref{appendix-F}, we obtain
\begin{equation}
F (p) \overset{p \to \infty}{\longrightarrow}
\lb\begin{array}{c@{\quad}l} 
c_F \, p^{D-4} + \frac{2 \kappa_2}{p^2}& (2 < D < 4)\\
- \frac{1}{(4\pi)^2} \ln |p| + \textrm{const} + \frac{2
  \kappa_2}{p^2}& (D=4)\,,
\end{array}\right.
\label{cF-asymptotic}
\end{equation}
where $c_F$ is given by (\ref{cF}).  Hence, we obtain
\begin{equation}
\op{\frac{1}{2}\op{\phi^2 (p)} \frac{1}{2}\op{\phi^2 (q)}}
\overset{|p-q| \to
  \infty}{\underset{p+q\,\mathrm{fixed}}{\longrightarrow}}
c_F\, p^{D-4} \, \delta (p+q) + \frac{2}{p^2} \frac{1}{2} \op{\phi^2 (p+q)}
\end{equation}
for $2 < D < 4$, and
\begin{eqnarray}
&&\op{\frac{1}{2}\op{\phi^2 (p)} \frac{1}{2}\op{\phi^2 (q)}}\nn\\
&&\quad\overset{|p-q| \to
  \infty}{\underset{p+q\,\mathrm{fixed}}{\longrightarrow}}
\left( - \frac{1}{(4\pi)^2} \ln |p| + \textrm{const} \right)
\delta (p+q) + \frac{2}{p^2} \frac{1}{2} \op{\phi^2 (p+q)}
\end{eqnarray}
for $D=4$.

\subsection*{Example 2: $\left[\left[\frac{1}{4!} \phi^4
      (p)\right]\left[\frac{1}{2} \phi^2 (q)\right]\right]$} 

The scale dimension of the product is $- y = D-6$.  Hence, in $D=4$,
the product mixes with $\frac{1}{2} \op{\phi^2 (p+q)}$.
Let
\begin{equation}
\op{ \frac{1}{4!} \op{\phi^4 (p)} \frac{1}{2} \op{\phi^2 (q)} }
= \frac{1}{4!} \op{\phi^4 (p)} \frac{1}{2} \op{\phi^2 (q)} + \PP_{42} (p,q)\,.
\end{equation}
Solving
\begin{eqnarray}
&&\left( 6 - D + p \cdot \partial_p + q \cdot \partial_q - \mathcal{D}
\right) \PP_{42} (p,q)\nn\\
&&= \int_r f(r) \frac{\delta}{\delta \Phi (r)} \frac{1}{4!} \op{\phi^4
  (p)} \frac{\delta}{\delta \Phi (-r)}  \frac{1}{2} \op{\phi^2
  (q)}\nn\\
&&= \frac{1}{4!} \int_{p_1, \cdots, p_4} \Phi (p_1) \cdots \Phi
(p_4)\, \delta (p_1 + \cdots + p_4 - (p+q)) \sum_{i=1}^4 f(p_i-q)\nn\\
&&\quad + \kappa_2 \frac{1}{2} \int_{p_1, p_2} \Phi (p_1) \Phi (p_2)
\delta (p_1+p_2-(p+q))\, \left( f(p_1-p) + f(p_1-q)\right)\,,
\end{eqnarray}
we obtain, for $2 < D < 4$,
\begin{eqnarray}
\PP_{42} (p,q) &=& \frac{1}{4!} \int_{p_1, \cdots, p_4} \Phi (p_1)
\cdots \Phi (p_4)\, \delta \left( p_1+\cdots + p_4 - (p+q)\right)
  \sum_{i=1}^4 h (p_i-q)\nn\\
&&+ \frac{1}{2} \int_{p_1, p_2} \Phi (p_1)\Phi (p_2) \delta
(p_1+p_2-(p+q)) \left( \kappa_2 \sum_{i=1,2} h (p_i-q) + F (q)
\right)\nn\\
&&+ \kappa_2 F (p) \delta (p+q)\,.
\end{eqnarray}
(See Fig. 2.)
\begin{figure}[t]
\centering
\includegraphics[width=10cm]{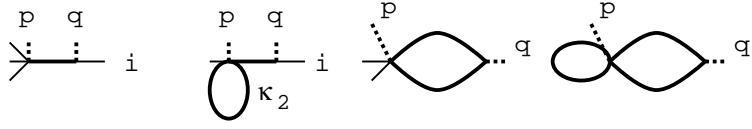}
\caption{Graphical representation of $\PP_{42} (p,q)$}
\end{figure}
The above expression is valid also for $D=4$ except that the ERG
equation for the product is modified to
\begin{equation}
\left(6-D+p \cdot \partial_p + q \cdot \partial_q - \mathcal{D}
\right) \op{ \frac{1}{4!} \op{\phi^4 (p)} \frac{1}{2} \op{\phi^2 (q)}
} = - \frac{1}{(4\pi)^2} \frac{1}{2} \op{\phi^2 (p+q)}\,.
\end{equation}

Using (\ref{cF-asymptotic}), we obtain
\begin{equation}
\op{\frac{1}{4!} \op{\phi^4 (p)} \frac{1}{2} \op{\phi^2 (q)}}
   \overset{|p-q|\to
   \infty}{\underset{p+q\,\mathrm{fixed}}{\longrightarrow}}
c_F \, p^{D-4} \frac{1}{2} \op{\phi^2 (p+q)} 
+ \frac{4}{p^2} \frac{1}{4!} \op{\phi^4 (p+q)}
\end{equation}
for $2 < D < 4$, and
\begin{equation}
\op{\frac{1}{4!} \op{\phi^4 (p)} \frac{1}{2} \op{\phi^2 (q)}}
   \overset{|p-q|\to
   \infty}{\underset{p+q\,\mathrm{fixed}}{\longrightarrow}}
\left( - \frac{1}{(4\pi)^2} \ln |p| + \textrm{const} \right)  \frac{1}{2} \op{\phi^2 (p+q)} 
+ \frac{4}{p^2} \frac{1}{4!} \op{\phi^4 (p+q)}
\end{equation}
for $D=4$.

\section{Multiple products\label{section-multiple}}

Multiple products of composite operators are defined just like the
product of two composite operators as single nonlocal composite
operators.  To start with, the product of three composite operators
$\Op_i$ of scale dimension $-y_i\,(i=1,2,3)$ is defined as a composite
operator of scale dimension $-(y_1+y_2+y_3)$ as
\begin{eqnarray}
&&\left[ \Op_1 (p) \Op_2 (q) \Op_3 (r) \right] \equiv \Op_1 (p) \Op_2
(q) \Op_3 (r)\nn\\
&&\qquad + \PP_{12} (p,q) \Op_3 (r) + \PP_{13} (p,r) \Op_2 (q) + \PP_{23}
(q,r) \Op_1 (p) + \PP_{123} (p,q,r)\,,
\end{eqnarray}
where the counterterm $\PP_{12}$ is the same counterterm that makes
\[
\op{\Op_1 (p) \Op_2 (q)} = \Op_1 (p) \Op_2 (q) + \PP_{12} (p,q)
\]
a composite operator.  The extra counterterm $\PP_{123} (p,q,r)$ has
to do with the three operators close to each other simultaneously.  We
obtain the ERG equation
\begin{eqnarray}
&&\left( \sum_{i=1}^3 y_i + p \cdot \partial_p + q \cdot \partial_q + r
  \cdot \partial_r - \mathcal{D} \right) \PP_{123} (p,q,r)\nn\\
&&= \int_s f (s)
   \cdot \left( \frac{\delta}{\delta \Phi(s)} \PP_{12} (p,q)
   \frac{\delta}{\delta \Phi(-s)} \Op_3 (r) \right.\nn\\
&&\quad \left.+ \frac{\delta}{\delta \Phi(s)} \PP_{23} (q,r)
   \frac{\delta}{\delta \Phi(-s)} \Op_1 (p) 
+ \frac{\delta}{\delta \Phi(s)} \PP_{31} (r,p)
   \frac{\delta}{\delta \Phi(-s)} \Op_2 (q)\right)\,.
\end{eqnarray}
If there is a local composite operator $\Op$ whose scale dimension
$-y$ satisfies
\begin{equation}
-y = -\sum_{i=1}^3 y_i - n\quad(n=0,1,\cdots)\,,
\end{equation}
the product $\op{\Op_1 (p) \Op_2 (q) \Op_3 (r)}$ may mix with $\Op
(p+q+r)$, and we obtain
\begin{equation}
\left( \sum_{i=1}^3 y_i + p \cdot \partial_p + q \cdot \partial_q + r
  \cdot \partial_r - \mathcal{D} \right) \op{\Op_1 (p) \Op_2 (q) \Op_3
  (r)} = d (p-q, q-r) \Op (p+q+r)\,,
\end{equation}
where $d (p-q,q-r)$ is a degree $n$ polynomial of $p-q, q-r$.

Proceeding further, we can define the product of four composite
operators as
\begin{eqnarray}
&&\left[ \Op_1 (p) \Op_2 (q) \Op_3 (r) \Op_4 (s)\right] \equiv \Op_1 (p) \Op_2
(q) \Op_3 (r) \Op_4 (s) \nn\\
&&\qquad + \PP_{12} (p,q) \Op_3 (r) \Op_4 (s) + \textrm{5 more terms} \nn\\
&&\qquad + \PP_{12} (p,q) \PP_{34} (r,s) + \PP_{13} (p,r) \PP_{24} (q,s) +
\PP_{14} (p,s) \PP_{23} (q,r)\nn\\
&&\qquad + \PP_{123} (p,q,r)\Op_4 (s) + \textrm{3 more terms}
 + \PP_{1234} (p,q,r,s)\,.
\end{eqnarray}
In the absence of mixing the last counterterm satisfies
\begin{eqnarray}
&&\left( \sum_{i=1}^4 y_i + p \cdot \partial_p + \cdots + s
  \cdot \partial_s - \mathcal{D} \right) \PP_{1234} (p,q,r,s)\nn\\
&&= \int_t f(t) \left( \frac{\delta \PP_{12} (p,q)}{\delta \Phi (t)}
  \frac{\delta \PP_{34} (r,s)}{\delta \Phi (-t)} + \cdots
+ \frac{\delta \PP_{123} (p,q,r)}{\delta \Phi (t)} \frac{\delta \Op_4
  (s)}{\delta \Phi (-t)} + \cdots \right)\,.
\end{eqnarray}
This can be generalized to higher order products of composite
operators.

Let $\Op (p)$ be a composite operator with scale dimension $- y < 0$.
(If it is a scalar, it is a relevant operator.)  The $n$-th order
product has scale dimension $- n y$. We can introduce a source $\JJ
(p)$ so that
\begin{equation}
\left[ \Op (p_1) \cdots \Op (p_n) \right]
= \frac{\delta^n}{\delta \JJ (-p_1) \cdots \delta \JJ
  (-p_n)} e^{\mathcal{W} [\JJ]}\Big|_{\JJ = 0}\,,
\end{equation}
where
\begin{eqnarray}
\mathcal{W} [\JJ] &=& \int_p \JJ (-p) \Op (p) +
\frac{1}{2} \int_{p_1, p_2} \JJ (-p_1) \JJ (-p_2)
\PP_2 (p_1, p_2)  \nn\\
&& + \frac{1}{3!} \int_{p_1, p_2, p_3} \JJ
(-p_1) \JJ (-p_2) \JJ (-p_3) \PP_3 (p_1, p_2, p_3) +
\cdots\,.
\end{eqnarray}
If no local composite operator has scale dimension as low as $- 2 y$,
there is no mixing between multiple products and local composite
operators.  (Since $\delta (p)$ has the lowest scale dimension $-D$,
there is no mixing if $- 2 y < - D$.)  In the absence of mixing,
$\mathcal{W} [\JJ]$ satisfies the ERG equation:
\begin{eqnarray}
&&\int_p \JJ(-p)\left(y + p \cdot \partial_p\right)
\frac{\delta}{\delta \JJ (-p)} 
  e^{\mathcal{W} [\JJ]}\nn\\
&&= \int_p J(-p) \left(- p \cdot \partial_p - \frac{D+2}{2} + \gamma
   \right) \frac{\delta}{\delta J(-p)} e^{\mathcal{W}
   [\JJ]}\nn\\
&&\quad + \int_p \left( - p \cdot \partial_p + 2 - 2 \gamma \right)
   R(p) \cdot \left(\frac{\delta W[J]}{\delta J (-p)}
   \frac{\delta}{\delta J(p)} + \frac{1}{2} \frac{\delta^2}{\delta
   J(p) \delta J(-p)} \right)  e^{\mathcal{W} [\JJ]}\,.\label{ERGeq-calW}
\end{eqnarray}

As the simplest example, consider $\Op (p) = \Phi (p)$ with $-y = -
\frac{D+2}{2} + \gamma$.
Eq.~(\ref{multi-Phi}) implies\cite{Rosten:2010vm}
\begin{equation}
\mathcal{W} [\JJ] = W \left[ J + \JJ \right] - W [J]\,.
\end{equation}
Another simple example is $\Op (p) = \frac{1}{2} \left[\phi^2
  (p)\right]$ with $-y=-2$ at the Gaussian fixed point ($2 < D \le
4$).  $\PP_2$ is $\PP_{22}$, obtained in Example 1 of
Sec.~\ref{section-examples}.  $\PP_n\, (n \ge 3)$ are given in the
form
\begin{eqnarray}
\PP_n (p_1, \cdots, p_n) &=& \frac{1}{2} \int_{p,q} \Phi (p) \Phi (q)\,
                             \delta \left( p+q-\sum_{i=1}^n
                             p_i\right)\, u_n (p_1, \cdots, p_n; p,
                             q)\nn\\
&& + v_n (p_1, \cdots, p_n) \,\delta \left( \sum_{i=1}^n
                             p_i\right)\,,
\end{eqnarray}
where $u_n, v_n$ are expressed graphically in Fig.~3.
\begin{figure}[t]
\centering
\includegraphics[width=8cm]{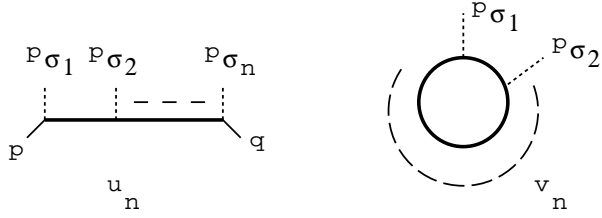}
\caption{Graphical representation of $u_n$ and $v_n$ --- $\sigma$ is a
permutation of $1,\cdots,n$}
\end{figure}
For example, we obtain
\begin{eqnarray}
u_3 (p_1,\cdots, p_3; p,q) &=& h (p-p_1) \left( h(p - p_1 - p_2) + h
                               (p-p_1-p_3)\right) + \cdots\,,\\
v_3 (p_1, \cdots, p_3) &=& \int_p h (p) h (p-p_1) h (p-p_1-p_2)\,.
\end{eqnarray}
At $D=4$, $\PP_2 (p_1,p_2)$ mixes with $\delta (p_1+p_2)$, and we must
change the left-hand side of (\ref{ERGeq-calW}) to
\begin{equation}
\left(\int_p \JJ (-p) \left(2 + p \cdot \partial_p\right)
\frac{\delta }{\delta \JJ (-p)} +
\frac{1}{(4\pi)^2} \frac{1}{2} \int_p \JJ (p) \JJ (-p)
\right) e^{\mathcal{W}[\JJ]}\,.
\end{equation}

\section{Away from a fixed point\label{section-massive}}

Let $g$ be a parameter with scale dimension $y_E > 0$.  The Wilson action
is parametrized by $g$.  Assuming the anomalous dimension is
independent of $\gamma$, we obtain the ERG equation of the action as
\begin{eqnarray}
y_E \, g \frac{\partial}{\partial g} e^{S(g) [\phi]} &=& 
\int_p \left( - p \cdot \partial_p \ln K(p) + \frac{D+2}{2} - \gamma +
 p \cdot \partial_p \right) \phi (p) \cdot \frac{\delta}{\delta \phi
                                                (p)} e^{S(g) [\phi]}\nn\\
&& + \int_p \left( - p \cdot \partial_p \ln R(p) + 2 - 2
   \gamma \right) \frac{K(p)^2}{R(p)} \frac{1}{2} \frac{\delta^2}{\delta
   \phi (p) \delta \phi (-p)} e^{S(g) [\phi]}\,.
\end{eqnarray}
The modified correlation functions defined by
\begin{equation}
\vvev{\phi (p_1) \cdots \phi (p_n)}_g \equiv
\prod_{i=1}^n \frac{1}{K(p_i)} \cdot \vev{\exp \left( - \int_p \frac{K(p)^2}
      {R(p)} \frac{1}{2} \frac{\delta^2}{\delta \phi (p) \delta
      \phi (-p)} \right)\, \phi (p_1) \cdots \phi (p_n)}_{S (g)}
\end{equation}
satisfy the scaling law:
\begin{equation}
\vvev{\phi (p_1 e^t) \cdots \phi (p_n e^t)}_{g\, e^{y_E t}}
= \exp \left( n t \left(- \frac{D+2}{2} + \gamma \right) \right)
\vvev{\phi (p_1) \cdots \phi (p_n)}_g\,.
\end{equation}

Rewriting the ERG equation for
\begin{subequations}
\begin{eqnarray}
W (g) [J] &\equiv& \frac{1}{2} \int_p \frac{J (p) J(-p)}{R(p)} + S (g)
[\phi]\,,\\
J (p) &\equiv& \frac{R(p)}{K(p)} \phi (p)\,,
\end{eqnarray}
\end{subequations}
we obtain
\begin{eqnarray}
y_E g \frac{\partial}{\partial g} e^{W (g)[J]} 
&=& \int_p J(-p) \left( - p \cdot \partial_p - \frac{D+2}{2} + \gamma
      \right) \frac{\delta}{\delta J(-p)} e^{W (g)[J]}\nn\\
&& \quad +  \int_p \left( - p \cdot \partial_p + 2 - 2 \gamma \right)
   R (p) \cdot \frac{1}{2} \frac{\delta^2}{\delta J(p) \delta J(-p)}
   e^{W (g)[J]}\,.\label{ERGeq-W-away}
\end{eqnarray}
Comparing this with (\ref{ERGeq-calW}), we find that for the constant
source
\begin{equation}
\JJ (p) = g \, \delta (p)\,,
\end{equation}
the sum
\begin{equation}
W(g) [J] \equiv W[J] + \mathcal{W} [\JJ]
\end{equation}
satisfies (\ref{ERGeq-W-away}).  Therefore, 
\begin{equation}
e^{- W[J]} \frac{\partial^n}{\partial g^n} e^{W (g) [J]}\Big|_{g=0} =
\op{\mathcal{Q} (0) \cdots \mathcal{Q} (0)}
\end{equation}
is the $n$-th order product of the zero momentum composite operator
\[
\mathcal{Q} (0) \equiv \frac{\partial W (g)[J]}{\partial g}\Big|_{g=0}
\]
of scale dimension $- y_E$, defined at the fixed point.

Considering operator products, it is even more convenient to
introduce the effective action with an infrared cutoff:
\begin{subequations}
\begin{eqnarray}
\Gamma (g) [\Phi] &\equiv& \frac{1}{2} \int_p R(p) \Phi (p) \Phi (-p)
+ W (g) [J] - \int_p J(-p) \Phi (p)\,,\\
\Phi (p) &\equiv& \frac{\delta W(g) [J]}{\delta J(-p)}\,.
\end{eqnarray}
\end{subequations}
Then, a composite operator of scale dimension $-y$ satisfies
\begin{equation}
\left( y + y_E \,g \partial_g + p \cdot \partial_p - \mathcal{D} \right)
\Op (p) = 0\,, 
\end{equation}
where $\mathcal{D}$ is given by (\ref{calD-def}) except that $G$ is
now defined with  the $g$-dependent $W$ as
\begin{equation}
G(g)_{p,q} [\Phi] \equiv \frac{\delta^2 W(g) [J]}{\delta J(p) \delta
  J(q)}\,.
\end{equation}
The discussion of Sec.~\ref{section-product} goes through as long as
we use the $g$-dependent $\mathcal{D}$.  We introduce $g$-dependent
coefficients $c_{12,i} (g; p)$ and $d_{12,i} (g;p)$ via
(\ref{expand-PP}) and (\ref{ERG-PP-RHS}), respectively.
Eq.~(\ref{diffeq-c}) is replaced by
\begin{equation}
\left( p \cdot \partial_p + y_E\, g
  \frac{\partial}{\partial g} + y_1 + y_2 - y_i \right) c_{12,i} (g;
p) = d_{12,i} (g; p)\,.
\end{equation}
Since the locality of the cutoff function $R$ gives
\begin{equation}
d_{12,i} (g; p) \overset{p \to \infty}{\longrightarrow} 0\,,
\end{equation}
we obtain the asymptotic behavior
\begin{equation}
c_{12,i} (g; p-q) \overset{|p-q| \to
  \infty}{\underset{p+q\,\mathrm{fixed}}{\longrightarrow}}
p^{-y_1-y_2+y_i} \left( C_{12,i} + \frac{g}{p^{y_E}}\, C'_{12,i} +
\mathrm{O} \left(\frac{g^2}{p^{2 y_E}}\right) \right)\,,
\end{equation}
assuming the analyticity of $c_{12,i} (g;p)$ in $g$.

The simplest example is given by the massive Gaussian theory
\begin{equation}
\Gamma (m^2) [\Phi] = - \frac{1}{2} \int_p (p^2 + m^2) \Phi (p) \Phi
(-p)\,,
\end{equation}
where the squared mass $m^2$ plays the role of $g$. (See Appendix
\ref{appendix-Gaussian} for the construction of composite operators.)
We can show, for $3 < D < 4$ (the term proportional to $m^2$ is not
singular for $2 < D \le 3$)
\begin{equation}
\op{\frac{1}{2} \op{\phi^2 (p)} \frac{1}{2} \op{\phi^2 (q)}}
\overset{|p-q| \to
  \infty}{\underset{p+q\,\mathrm{fixed}}{\longrightarrow}}
c_F \left( 1 + 2(D-3) \frac{m^2}{p^2} \right) p^{D-4} \, \delta (p+q)
 + \frac{2}{p^2}\, \frac{1}{2} \op{\phi^2 (p+q)}\,,
\end{equation}
and for $D=4$
\begin{eqnarray}
\op{\frac{1}{2} \op{\phi^2 (p)} \frac{1}{2} \op{\phi^2 (q)}}
&\overset{|p-q| \to
  \infty}{\underset{p+q\,\mathrm{fixed}}{\longrightarrow}}&
\left( - \frac{1}{(4\pi)^2} \left(1 + \frac{2 m^2}{p^2}\right) \ln |p| + \mathrm{const} \right)
 \delta (p+q) \nn\\
&&\quad + \frac{2}{p^2}\, \frac{1}{2} \op{\phi^2 (p+q)}\,.
\end{eqnarray}

\section{Conclusions\label{section-conclusions}}

In this work we have studied the OPE (operator product expansions) in
the framework of ERG (the exact renormalization group).  The key
concepts underlying our analysis are the composite operators and their
products, which we define respectively in Sec.~\ref{section-comp} and
Sec.~\ref{section-product}.  We have argued that the ERG differential
equation associated with the product of two operators can be expanded
in a local basis of composite operators, leading to the ERG
differential equations for the Wilson OPE coefficients.  Particular
attention has been paid to the form of the equation at a fixed point.
It is important to stress that the Wilson coefficients are defined in
the large momentum limit, i.e., $p \rightarrow \infty$. Taking this
limit allows us to eliminate spurious contributions dependent on the
cutoff.

We have tested our method by considering some explicit examples in
Sec.~\ref{section-examples}.  At the technical level, we have found it
convenient to first solve the ERG differential equation by taking into
account the analyticity of the equation at the zero momentum before
taking the large momentum limit.  In Sec.~\ref{section-multiple} we
have generalized our discussion to include the definition of multiple
products of a composite operator, and in Sec.~\ref{section-massive} we
have considered the ERG differential equations away from the fixed
point.

Although the examples we have given are for the Gaussian fixed point,
we would like to stress that the ERG differential equations discussed
in Sec.~\ref{section-product} are nonperturbative and can be employed for
nonperturbative, albeit approximate, computations.  In this sense
suitable approximation schemes should be devised.  Nonperturbative
approximation schemes, such as the BMW \cite{Blaizot:2005wd}, may be
employed to solve the ERG differential equations for the Wilson
coefficients.

Note added: after completion of the present work, we learned that
Prof.~H.~Osborn had similar ideas as ours about the products of
composite operators. (Sec.~3.3 of \cite{Osborn:2015})

\appendix

\section{More background on ERG\label{appendix-ERG}}

To make the content of Sec.~\ref{section-comp} easier to understand
for the readers already familiar with \cite{Polchinski:1983gv} but not
with \cite{Wilson:1973jj} (and the subsequent extensions done more
recently), we would like to summarize the basics of the ERG formalism
adopted in this paper.  We rely on perturbation theory for intuition.

In \cite{Polchinski:1983gv} the Wilson action is given in the form
\begin{equation}
S_\Lambda [\phi] = - \frac{1}{2} \int_p \frac{p^2}{K(p/\Lambda)} \phi (p)
\phi (-p) + S_{\Lambda, I} [\phi]\,,
\end{equation}
where the first term gives the propagator
\[
\frac{K(p/\Lambda)}{p^2}
\]
that damps rapidly for $p > \Lambda$, and the second term gives
interactions.  To preserve physics below the momentum scale $\Lambda$,
the interaction part must obey the following differential
equation\cite{Polchinski:1983gv}:
\begin{equation}
- \Lambda \frac{\partial}{\partial \Lambda} S_{\Lambda, I} [\phi]
=\int_p \frac{\Lambda \frac{\partial
    K(p/\Lambda)}{\partial \Lambda}}{p^2}  \frac{1}{2} \lb
\frac{\delta S_{\Lambda, 
    I} [\phi]}{\delta \phi (-p)} \frac{\delta S_{\Lambda,
    I} [\phi]}{\delta \phi (p)} + \frac{\delta^2 S_{\Lambda,
    I} [\phi]}{\delta \phi (-p) \delta \phi (p)} \rb\,.\label{polchinski-I}
\end{equation}
We can rewrite this equation for the whole action as
\begin{eqnarray}
- \Lambda \frac{\partial}{\partial \Lambda} S_\Lambda [\phi]
&=& \int_p \Lambda \frac{\partial}{\partial \Lambda} \ln
  K(p/\Lambda) \cdot \phi (p) \frac{\delta}{\delta \phi (p)} S_\Lambda
  [\phi]\nn\\
&& + \int_p \frac{\Lambda \frac{\partial K(p/\Lambda)}{\partial
    \Lambda}}{p^2} \frac{1}{2} \lb \frac{\delta S_\Lambda}{\delta \phi
  (p)} \frac{\delta S_\Lambda}{\delta \phi (-p)} + \frac{\delta^2
  S_\Lambda}{\delta \phi (p) \delta \phi (-p)} \rb\,.\label{polchinski-full}
\end{eqnarray}

The correlation functions calculated with $S_\Lambda$
\begin{equation}
\vev{\phi (p_1) \cdots \phi (p_n)}_\Lambda \equiv
\int [d\phi] e^{S_\Lambda [\phi]}\, \phi (p_1) \cdots \phi (p_n)
\end{equation}
are not entirely independent of $\Lambda$.  Take the sum of
diagrams contributing to the \textbf{connected} part of the $n$-point
function for $n > 2$. (Fig.~4)
\begin{figure}[t]
\centering
\includegraphics{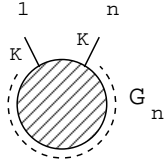}
\caption{Graph for the connected $n$-point function ($n > 2$)}
\end{figure}
Polchinski's equation (\ref{polchinski-I}) guarantees that the shaded
blob (denoted $G_n$) of Fig.~4 is independent of $\Lambda$.  But the
external propagators are multiplied by $K$, and we obtain
\begin{equation}
\vev{\phi (p_1) \cdots \phi (p_n)}^{\mathrm{connected}}_\Lambda
= G_n (p_1,\cdots, p_n) \prod_{i=1}^n \frac{K(p_i/\Lambda)}{p_i^2}\,.
\end{equation}
For small momenta $p_i$, the cutoff function $K(p_i/\Lambda)$ is
almost $1$.  We only need to divide the correlation function by a
product of $K$'s to make this strictly $\Lambda$-independent:
\begin{equation}
\prod_{i=1}^n \frac{1}{K(p_i/\Lambda)}\, \vev{\phi (p_1) \cdots \phi
  (p_n)}^{\mathrm{connected}}_\Lambda\,.
\end{equation}

We are left with the two-point function which is given by
\begin{equation}
\vev{\phi (p) \phi (q)}_\Lambda
= \left[ \frac{K(p/\Lambda)}{p^2}+ \frac{K(p/\Lambda)}{p^2} G_2
(p) \frac{K(p/\Lambda)}{p^2} \right] \delta (p+q)\,,
\end{equation}
where $G (p)$ is independent of $\Lambda$.  (Fig.~5)
\begin{figure}[t]
\centering
\includegraphics{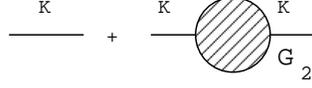}
\caption{Two-point function}
\end{figure}
Again, this is almost independent of $\Lambda$ for $p < \Lambda$.  To
make the two-point function strictly $\Lambda$-independent, we first
modify it by subtracting a high momentum propagator
\begin{equation}
\frac{K(p/\Lambda)\left(1 - K(p/\Lambda)\right)}{p^2}
\label{highmomentum-prop}
\end{equation}
and then divide the result by $K(p/\Lambda)^2$:
\begin{eqnarray}
&&\frac{1}{K(p/\Lambda)^2}
\left( \vev{\phi (p) \phi (q)}_\Lambda - \frac{K(p/\Lambda)\left(1 -
      K(p/\Lambda)\right)}{p^2} \delta (p+q) \right)\nn\\
&&= \left(\frac{1}{p^2} + G_2 (p) \right) \delta (p+q)\,.
\end{eqnarray}
This is independent of $\Lambda$.

We have thus explained that $\Lambda$-independent connected
correlation functions are given by
\begin{subequations}
\begin{eqnarray}
&&\vvev{\phi (p) \phi (q)} \equiv
 \frac{1}{K(p/\Lambda)^2} \left( \vev{\phi (p)\phi (q)}_\Lambda -
  \frac{K(p/\Lambda)\left(1 - K(p/\Lambda)\right)}{p^2} \delta (p+q)
\right)\,,\\
&&\vvev{\phi (p_1) \cdots \phi (p_n)}^{\mathrm{connected}} \equiv
\prod_{i=1}^n \frac{1}{K(p_i/\Lambda)}\,\times \vev{\phi (p_1) \cdots \phi
  (p_n)}^{\mathrm{connected}}_\Lambda\,,
\end{eqnarray}
\end{subequations}
where $n > 2$.  Including the disconnected parts, the
$\Lambda$-independent four-point correlation function is given by
\begin{eqnarray}
\vvev{\phi (p_1) \cdots \phi (p_4)} &\equiv&
\vvev{\phi (p_1) \cdots \phi (p_4)}^{\mathrm{connected}}
 + \vvev{\phi (p_1) \phi (p_2)} \vvev{\phi (p_3) \phi (p_4)}
+ (\textrm{t, u-channels})\nn\\
&=& \prod_{i=1}^4 \frac{1}{K(p_i/\Lambda)} \Big[
\vev{\phi (p_1) \cdots \phi
  (p_4)}_\Lambda^{\mathrm{connected}}\nn\\
&& \quad - \frac{K(p_1/\Lambda)\left(1 - K(p_1/\Lambda)\right)}{p_1^2}
\delta (p_1+p_2) \, \vev{\phi (p_3) \phi (p_4)}_\Lambda\nn\\
&& \quad - \frac{K(p_3/\Lambda)\left(1 - K(p_3/\Lambda)\right)}{p_3^2}
\delta (p_3+p_4) \, \vev{\phi (p_1) \phi (p_2)}_\Lambda \nn\\
&& \quad+  \frac{K(p_1/\Lambda)\left(1 - K(p_1/\Lambda)\right)}{p_1^2}
\delta (p_1+p_2)\frac{K(p_3/\Lambda)\left(1 - K(p_3/\Lambda)\right)}{p_3^2}
\delta (p_3+p_4) \nn\\
&&\quad + (\textrm{t, u-channels}) \Big]\,.
\end{eqnarray}
This structure generalizes to higher-point functions.  Using a formal
but more convenient notation, we can express $\Lambda$-independent
correlation functions by
\begin{eqnarray}
&&\vvev{\phi (p_1) \cdots \phi (p_n)}_\Lambda \equiv \prod_{i=1}^n
\frac{1}{K(p_i/\Lambda)}\nn\\
&& \qquad \times 
\vev{\exp \left( - \int_p
    \frac{K(p/\Lambda)\left(1-K(p/\Lambda)\right)}{p^2} \frac{1}{2}
    \frac{\delta^2}{\delta \phi (p)\delta \phi (-p)}\right)
\phi (p_1)\cdots \phi (p_n)}_\Lambda\,.
\end{eqnarray}

These $\Lambda$-independent correlation functions were first
introduced in \cite{Sonoda:2015bla} and termed \textbf{modified}
correlation functions.  Two Wilson actions are called there equivalent
if their modified correlation functions are the same.  We find the
word ``modified'' somewhat misleading since these are the proper
correlation functions given by the Wilson action.

The ERG differential equation (\ref{polchinski-full}) still differs
from the ERG equations of the main text.  In order to discuss a fixed
point of the renormalization group, we need to introduce an anomalous
dimension and adopt the dimensionless convention to fix the momentum
cutoff.  Let us explain this one by one.
\begin{enumerate}
\item \textbf{Anomalous dimension} --- To keep the kinetic term of
  $S_\Lambda$ independent of $\Lambda$, we must introduce an
  appropriate $\Lambda$-dependence to the normalization of $\phi$.
  This introduces an anomalous dimension $\gamma_\Lambda$ so that
\begin{equation}
\vvev{\phi (p_1) \cdots \phi (p_n)}_\Lambda
= \left(\frac{Z_\Lambda}{Z_{\Lambda'}}\right)^{\frac{n}{2}} \vvev{\phi
  (p_1) \cdots \phi (p_n)}_{\Lambda'} 
\end{equation}
where
\begin{equation}
- \Lambda \frac{\partial}{\partial \Lambda} \ln Z_\Lambda = 2
\gamma_\Lambda\,.
\end{equation}
To obtain this we must change (\ref{polchinski-full})
to
\begin{eqnarray}
- \Lambda \frac{\partial S_\Lambda}{\partial \Lambda} &=& \int_p
\left( \Lambda \frac{\partial}{\partial \Lambda} \ln K(p/\Lambda) -
  \gamma_\Lambda \right) \phi (p) \frac{\delta S_\Lambda}{\delta \phi
  (p)}\nn\\
&& + \int_p \frac{1}{p^2} \left( \Lambda \frac{\partial
    K(p/\Lambda)}{\partial \Lambda} - 2 \gamma_\Lambda K(p/\Lambda) \left(1 -
    K(p/\Lambda)\right)\right) \nn\\
&&\qquad\qquad \times \frac{1}{2} \lb \frac{\delta
    S_\Lambda}{\delta \phi (-p)} \frac{\delta S_\Lambda}{\delta \phi
    (p)} + \frac{\delta^2 S_\Lambda}{\delta \phi (p) \delta \phi (-p)}
  \rb\,.
\end{eqnarray}
(There are other ways of introducing $\gamma_\Lambda$ such as the one
given in \cite{Ball:1994ji}.  Here we have followed
\cite{Sonoda:2015bla,Igarashi:2016qdr}.)
Rewriting the second integral of the right-hand side, we obtain
\begin{eqnarray}
- \Lambda \frac{\partial S_\Lambda}{\partial \Lambda} &=& \int_p
\left( \Lambda \frac{\partial}{\partial \Lambda} \ln K(p/\Lambda) -
  \gamma_\Lambda \right) \phi (p) \frac{\delta S_\Lambda}{\delta \phi
  (p)}\nn\\
&& + \int_p \left( \Lambda \frac{\partial}{\partial \Lambda} \ln
  \frac{K(p/\Lambda)}{1-K(p/\Lambda)} - 2 \gamma_\Lambda \right) 
\frac{K(p/\Lambda)\left(1-K(p/\Lambda)\right)}{p^2}\nn\\
&&\qquad\qquad \times \frac{1}{2}
 \lb \frac{\delta
    S_\Lambda}{\delta \phi (-p)} \frac{\delta S_\Lambda}{\delta \phi
    (p)} + \frac{\delta^2 S_\Lambda}{\delta \phi (p) \delta \phi (-p)}
  \rb\,.
\end{eqnarray}
\item \textbf{Dimensionless convention} --- Finally, to obtain a fixed
  point we must adopt the dimensionless convention by measuring
  physical quantities in powers of appropriate powers of the cutoff
  $\Lambda$.  This serves the purpose of rescaling, fixing the
  momentum cutoff at an arbitrary but fixed scale.  We make the
  following replacement:
\begin{equation}
\begin{array}{r@{~\longrightarrow~}l}
\frac{\Lambda}{\mu} & e^{-t}\\
\gamma_\Lambda & \gamma_t\\
\frac{p}{\Lambda} & p\\
\Lambda^{\frac{D+2}{2}} \phi (p) & \phi (p)\\
S_\Lambda & S_t
\end{array}
\end{equation}
where $\mu$ is an arbitrary momentum scale corresponding to the origin
$t=0$ of the logarithmic momentum scale.  The ERG equation becomes
\begin{eqnarray}
  \partial_t S_t [\phi] &=& \int_p \left( - p \cdot \partial_p \ln K(p)
    + \frac{D+2}{2} - \gamma_t + p \cdot \partial_p \right) \phi (p) \cdot
  \frac{\delta S_t [\phi]}{\delta \phi (p)}\nn\\
  && + \int_p \frac{1}{p^2} \left( - p \cdot \partial_p \ln
    \frac{K(p)}{1-K(p)} - 2 \gamma_t \right)  
  \frac{K(p)\left(1-K(p)\right)}{p^2} \nn\\
  &&\qquad\qquad \times \frac{1}{2} \lb
  \frac{\delta S_t [\phi]}{\delta \phi (-p)} \frac{\delta S_t
    [\phi]}{\delta \phi (p)} + \frac{\delta^2 S_t [\phi]}{\delta \phi
    (p) \delta \phi (-p)}\rb\,.
\end{eqnarray}
Introducing
\begin{equation}
R (p) \equiv p^2 \frac{K(p)}{1-K(p)}
\end{equation}
we can rewrite the above as
\begin{eqnarray}
  \partial_t S_t [\phi] &=& \int_p \left( - p \cdot \partial_p \ln K(p)
    + \frac{D+2}{2} - \gamma_t + p \cdot \partial_p \right) \phi (p) \cdot
  \frac{\delta S_t [\phi]}{\delta \phi (p)}\nn\\
  && + \int_p \frac{1}{p^2} \left( - p \cdot \partial_p \ln R(p) + 2
     - 2 \gamma_t \right) \frac{K(p)^2}{R(p)}  \nn\\
  &&\qquad\qquad \times \frac{1}{2} \lb
  \frac{\delta S_t [\phi]}{\delta \phi (-p)} \frac{\delta S_t
    [\phi]}{\delta \phi (p)} + \frac{\delta^2 S_t [\phi]}{\delta \phi
    (p) \delta \phi (-p)}\rb\,.
\end{eqnarray}
At the fixed point, the left-hand side vanishes, and we obtain
(\ref{ERG-S}) at the beginning of Sec.~\ref{section-comp}.
Though the cutoff function $R$ is given in terms of $K$ in the above
summary, we can take $K$ and $R$ independently as long as they satisfy
the conditions listed in  Sec.~\ref{section-comp}.\cite{Sonoda:2015bla}
\end{enumerate}

\section{Composite operators for the massive free theory\label{appendix-Gaussian}}

We consider the massive free theory in $D > 2$:
\begin{equation}
\Gamma [\Phi] = - \frac{1}{2} \int_p (p^2+m^2) \Phi (p) \Phi (-p)\,.
\end{equation}
The high-momentum propagator is given by
\begin{equation}
G_{p,q} [\Phi] \equiv \frac{\delta^2 \Gamma [\Phi]}{\delta \Phi (p)
  \delta \Phi (q)} = h (m^2, p)  \delta (p+q)\,,
\end{equation}
where
\begin{equation}
h (m^2, p) \equiv \frac{1}{p^2 + m^2 + R(p)}\,.
\end{equation}
We define
\begin{equation}
f(m^2, p) \equiv \left(2 + 2 m^2 \partial_{m^2} + p
  \cdot \partial_p\right) h (m^2, p) = \frac{(2 - p \cdot \partial_p)
  R(p)}{\left(p^2 + m^2 + R(p)\right)^2}\,.
\end{equation}

A composite operator of scale dimension $(-y)$ satisfies
\begin{equation}
\left(y + 2 m^2 \frac{\partial}{\partial m^2} + p \cdot \partial_p  -
  \mathcal{D} \right) \Op (p) 
= 0\,,\label{ERG-Op-massive}
\end{equation}
where $\mathcal{D}$ is given by
\begin{equation}
\mathcal{D} = \int_q \lb \left( q \cdot \partial_q + \frac{D+2}{2}
\right) \Phi (q) \cdot \frac{\delta}{\delta \Phi (q)} + f(m^2, q)
\frac{1}{2} \frac{\delta^2}{\delta \Phi (q)\delta \Phi (-q)} \rb\,.
\end{equation}

A generic (even) scalar composite operator is written as
\begin{equation}
\Op (p) = \sum_{n=0}^N \frac{1}{(2n)!} \int_{p_1, \cdots, p_{2n}} \Phi
            (p_1) \cdots \Phi (p_{2n})\,\delta \left(\sum_{i=1}^{2n} p_i - p
            \right)\, O_{2n} (p_1, \cdots, p_{2n})\,.
\end{equation}
Substituting this into (\ref{ERG-Op-massive}), we obtain
\begin{subequations}
\begin{eqnarray}
&&\left( 2 m^2 \partial_{m^2} + \sum_{i=1}^{2N} p_i
   \cdot \partial_{p_i} + y + N (D-2) - D \right) O_{2N} (p_1,
   \cdots, p_{2N}) = 0\\
&&\left( 2 m^2 \partial_{m^2} + \sum_{i=1}^{2n} p_i
   \cdot \partial_{p_i} + y+ n(D-2) - D \right) O_{2n} (p_1,
   \cdots, p_{2n}) \nn\\
&&\qquad\qquad = \frac{1}{2} \int_q f(m^2, q)\, O_{2(n+1)} (p_1,
   \cdots, p_{2n}, q, -q)\,.\quad (n \le N-1)
\end{eqnarray}
\end{subequations}

The operator with the lowest scale dimension is the identity operator
$\delta (p)$ with $y = D$.  The second lowest dimensional operator, with
$y = 2$, is given by
\begin{equation}
\frac{1}{2} \op{\phi^2 (p)} = \frac{1}{2} \int_{p_1, p_2} \Phi (p_1)
\Phi (p_2)\, \delta (p_1+p_2-p)
 + \kappa_2 (m^2) \, \delta (p)\,,\label{appendix-Gaussian-phi2}
\end{equation}
where $\kappa_2 (m^2)$ satisfies
\begin{equation}
\left( 2 - D + 2 m^2 \frac{d}{dm^2}\right) \kappa_2 (m^2) = \frac{1}{2}
\int_p f(m^2, p)\,.\label{kappa2-diffeq}
\end{equation}
We expect $\kappa_2 (m^2)$ to be analytic at $m^2=0$ from the locality
of the composite operator $\frac{1}{2} \op{\phi^2 (p)}$; any
non-analytic behavior should result from the integration over the
momentum modes below the cutoff.  Now, Eq.~(\ref{kappa2-diffeq}) has a
homogeneous solution proportional to $(m^2)^{\frac{D-2}{2}}$.  For $2
< D < 4$, this is not analytic, and the equation has a unique analytic
solution satisfying
\begin{equation}
\kappa_2 (0) = \frac{1}{2-D} \frac{1}{2} \int_p f(0, p)\,.
\end{equation}
For $D=4$ (and higher even dimensions), we have a problem:
(\ref{kappa2-diffeq}) has no analytic solution because the right-hand
side has a term proportional to $m^2$.  We must subtract the linear
term and solve instead
\begin{subequations}
\label{kappa2-4D}
\begin{equation}
\left( - 2 + 2 m^2 \frac{d}{dm^2}\right) \kappa_2 (m^2) = \frac{1}{2}
\int_p f(m^2, p) - m^2 \frac{1}{2} \int_p \frac{\partial}{\partial m^2} f(m^2,
  p)\Big|_{m^2=0} \,,\label{kappa2-4D-diffeq}
\end{equation}
where
\begin{equation}
- \frac{1}{2} \int_p \frac{\partial}{\partial m^2} f(m^2,
  p)\Big|_{m^2=0} = \int_p f(0,p) h(0,p) = \frac{1}{(4\pi)^2}
\end{equation}
as calculated in (\ref{integral-fh}).  (\ref{kappa2-4D-diffeq}) has an
analytic solution, but the solution is not unique due to the analytic
homogeneous solution $m^2$.  This simply means that
$\frac{1}{2}\op{\phi^2} (p)$ mixes with $m^2 \delta (p)$ under
scaling.  We remove the ambiguity by adopting an arbitrary convention
such as
\begin{equation}
\frac{d}{dm^2} \kappa_2 (m^2)\Big|_{m^2=0} = 0\,.
\end{equation}
\end{subequations}
The operator thus defined satisfies
\begin{equation}
\left( 2 + 2 m^2 \frac{\partial}{\partial m^2} + p \cdot \partial_p -
  \mathcal{D} \right) \frac{1}{2} \op{\phi^2 (p)} =
\frac{m^2}{(4\pi)^2} \delta (p)\,,
\end{equation}
where the right-hand side implies mixing.  Any alternative choice
\begin{equation}
\frac{1}{2} \op{\phi^2 (p)} + \mathrm{const} \times m^2 \delta (p)
\end{equation}
is equally good as an element of a basis of composite operators.

The operator with $y = 4-D$ can be constructed similarly:
\begin{eqnarray}
\frac{1}{4!} \op{\phi^4 (p)} &=& \frac{1}{4!} \int_{p_1, \cdots, p_4}
\Phi (p_1) \cdots \Phi (p_4)\, \delta \left(\sum_{i=1}^4 p_i - p
\right)\nn\\
&& + \kappa_2 (m^2) \frac{1}{2} \int_{p_1, p_2} \Phi (p_1) \Phi
(p_2)\, \delta (p_1+p_2-p) + \kappa_4 (m^2)\, \delta (p)\,,
\end{eqnarray}
where $\kappa_4 (m^2)$ is defined by
\begin{equation}
\left( 2(2-D) + 2 m^2 \frac{d}{dm^2} \right) \kappa_4 (m^2) = \kappa_2
(m^2) \frac{1}{2} \int_p f(m^2, p)\,.
\label{kappa4-diffeq}
\end{equation}
For $2 < D < 4$, we obtain
\begin{equation}
\kappa_4 (m^2) = \frac{1}{2} \kappa_2 (m^2)^2\,.
\end{equation}
For $D=3$, the solution is ambiguous by a constant multiple of $m^2$.
(No subtraction is necessary; the right-hand side of
(\ref{kappa4-diffeq}) has no term linear in $m^2$.) For $D=4$, the
analyticity at $m^2=0$ demands that we modify the equation to
\begin{equation}
\left( - 4 + 2 m^2 \frac{d}{dm^2} \right) \kappa_4 (m^2) = \kappa_2
(m^2) \left( \frac{1}{2} \int_p f(m^2, p) +
  \frac{m^2}{(4\pi)^2}\right)\,,
\end{equation}
where $\kappa_2 (m^2)$ is determined by (\ref{kappa2-4D}).  The
solution is given by
\begin{equation}
\kappa_4 (m^2) = \frac{1}{2} \kappa_2 (m^2)^2 + \mathrm{const} \cdot
m^4\,.
\end{equation}
Again as a convention, we may impose
\begin{equation}
\left(\frac{d}{dm^2}\right)^2 \kappa_4 (m^2)\Big|_{m^2=0} = 0
\end{equation}
to fix the constant.  The operator thus defined satisfies
\begin{equation}
\left( 2 m^2 \frac{\partial}{\partial m^2} + p \cdot \partial_p -
  \mathcal{D} \right) \frac{1}{4!} \op{\phi^4 (p)} =
\frac{m^2}{(4\pi)^2} \frac{1}{2} \op{\phi^2 (p)}\,,
\end{equation}
which implies that $\frac{1}{4!} \op{\phi^4 (p)}$ mixes with $m^2
\frac{1}{2} \op{\phi^2 (p)}$ under scaling.

The operator $p^2 \frac{1}{2}\op{\phi^2 (p)}$ has $y=0$.  The other
operator with $y=0$ is the equation-of-motion composite
operator\cite{Becchi:1996an, Igarashi:2009tj, Rosten:2010vm} given by
\begin{eqnarray}
\mathcal{E} (p) &\equiv& - e^{-S} \int_q K(q) \frac{\delta}{\delta \phi
  (q)} \left( \Phi (q+p) e^S \right)\nn\\
&=& \frac{1}{2} \int_{p_1, p_2} \Phi (p_1) \Phi (p_2) \delta (p_1 +
p_2 - p) \, \sum_{i=1}^2 (p_i^2 + m^2) \nn\\
&&\quad - \int_q \frac{R(q)}{q^2+m^2+R(q)}\, \delta (p)\,.
\end{eqnarray}
For $2 < D \le 4$, all the other scalar operators have $y < 0$.

\section{Asymptotic behavior of $F(p)$\label{appendix-F}}

For $2 < D < 4$, 
\begin{equation}
F(p) \equiv \frac{1}{2} \int_q h(q) h (q+p)
\end{equation}
satisfies the differential equation
\begin{equation}
\left(4-D+p \cdot \partial_p \right) F(p) = \int_q f(q) h (q+p)\,.
\end{equation}
Since
\begin{equation}
\int_q f(q) h (q+p) \overset{p \to \infty}{\longrightarrow}
\frac{1}{p^2} \int_q f(q) \,,
\end{equation}
we obtain
\begin{equation}
F (p) \overset{p \to \infty}{\longrightarrow}
c_F\, p^{D-4} + \frac{1}{p^2} \frac{1}{2-D} \int_q f(q) = c_F\,
p^{D-4} + \frac{2 \kappa_2}{p^2}\,,
\end{equation}
where $c_F$ is a constant.  From
\begin{equation}
c_F \, p^{D-4} = \frac{1}{2} \int_q \frac{1}{q^2 (p+q)^2}
\end{equation}
we obtain
\begin{equation}
c_F =  \frac{1}{(4\pi)^{\frac{D}{2}}} \frac{\Gamma
    \left(2-\frac{D}{2}\right) \Gamma \left(\frac{D}{2}-1\right)^2}{2
    \Gamma (D-2)}\,.\label{cF}
\end{equation}

For $D=4$, $F(p)$ is determined by
\begin{equation}
p \cdot \partial_p F(p) = \int_q f(q) h (q+p) - \frac{1}{(4\pi)^2}
\end{equation}
and
\begin{equation}
F(0) = 0\,.
\end{equation}
For large $p$, the differential equation becomes
\begin{equation}
p \cdot \partial_p F (p) \overset{p \to \infty}{\longrightarrow} -
\frac{1}{(4 \pi)^2} + \frac{1}{p^2} \int_q f(q)\,,
\end{equation}
which gives
\begin{equation}
F (p) \overset{p \to \infty}{\longrightarrow} - \frac{1}{(4 \pi)^2} \ln |p| +
\textrm{const} + \frac{2 \kappa_2}{p^2}\,.
\end{equation}
The constant is determined by the initial condition $F(0)=0$, but it
depends on the choice of a cutoff function $R$.


\bibliography{paper}

\begin{thebibliography}{28}%
\makeatletter
\providecommand \@ifxundefined [1]{%
 \@ifx{#1\undefined}
}%
\providecommand \@ifnum [1]{%
 \ifnum #1\expandafter \@firstoftwo
 \else \expandafter \@secondoftwo
 \fi
}%
\providecommand \@ifx [1]{%
 \ifx #1\expandafter \@firstoftwo
 \else \expandafter \@secondoftwo
 \fi
}%
\providecommand \natexlab [1]{#1}%
\providecommand \enquote  [1]{``#1''}%
\providecommand \bibnamefont  [1]{#1}%
\providecommand \bibfnamefont [1]{#1}%
\providecommand \citenamefont [1]{#1}%
\providecommand \href@noop [0]{\@secondoftwo}%
\providecommand \href [0]{\begingroup \@sanitize@url \@href}%
\providecommand \@href[1]{\@@startlink{#1}\@@href}%
\providecommand \@@href[1]{\endgroup#1\@@endlink}%
\providecommand \@sanitize@url [0]{\catcode `\\12\catcode `\$12\catcode
  `\&12\catcode `\#12\catcode `\^12\catcode `\_12\catcode `\%12\relax}%
\providecommand \@@startlink[1]{}%
\providecommand \@@endlink[0]{}%
\providecommand \url  [0]{\begingroup\@sanitize@url \@url }%
\providecommand \@url [1]{\endgroup\@href {#1}{\urlprefix }}%
\providecommand \urlprefix  [0]{URL }%
\providecommand \Eprint [0]{\href }%
\providecommand \doibase [0]{http://dx.doi.org/}%
\providecommand \selectlanguage [0]{\@gobble}%
\providecommand \bibinfo  [0]{\@secondoftwo}%
\providecommand \bibfield  [0]{\@secondoftwo}%
\providecommand \translation [1]{[#1]}%
\providecommand \BibitemOpen [0]{}%
\providecommand \bibitemStop [0]{}%
\providecommand \bibitemNoStop [0]{.\EOS\space}%
\providecommand \EOS [0]{\spacefactor3000\relax}%
\providecommand \BibitemShut  [1]{\csname bibitem#1\endcsname}%
\let\auto@bib@innerbib\@empty
\bibitem [{\citenamefont {Becchi}(1996)}]{Becchi:1996an}%
  \BibitemOpen
  \bibfield  {author} {\bibinfo {author} {\bibfnamefont {C.}~\bibnamefont
  {Becchi}},\ }\bibfield  {title} {\enquote {\bibinfo {title} {{On the
  construction of renormalized gauge theories using renormalization group
  techniques}},}\ }\href@noop {} {\  (\bibinfo {year} {1996})},\ \Eprint
  {http://arxiv.org/abs/hep-th/9607188} {arXiv:hep-th/9607188 [hep-th]}
  \BibitemShut {NoStop}%
\bibitem [{\citenamefont {Pawlowski}(2007)}]{Pawlowski:2005xe}%
  \BibitemOpen
  \bibfield  {author} {\bibinfo {author} {\bibfnamefont {J.~M.}\ \bibnamefont
  {Pawlowski}},\ }\bibfield  {title} {\enquote {\bibinfo {title} {{Aspects of
  the functional renormalisation group}},}\ }\href {\doibase
  10.1016/j.aop.2007.01.007} {\bibfield  {journal} {\bibinfo  {journal} {Annals
  Phys.}\ }\textbf {\bibinfo {volume} {322}},\ \bibinfo {pages} {2831--2915}
  (\bibinfo {year} {2007})},\ \Eprint {http://arxiv.org/abs/hep-th/0512261}
  {arXiv:hep-th/0512261 [hep-th]} \BibitemShut {NoStop}%
\bibitem [{\citenamefont {Igarashi}\ \emph {et~al.}(2010)\citenamefont
  {Igarashi}, \citenamefont {Itoh},\ and\ \citenamefont
  {Sonoda}}]{Igarashi:2009tj}%
  \BibitemOpen
  \bibfield  {author} {\bibinfo {author} {\bibfnamefont {Y.}~\bibnamefont
  {Igarashi}}, \bibinfo {author} {\bibfnamefont {K.}~\bibnamefont {Itoh}}, \
  and\ \bibinfo {author} {\bibfnamefont {H.}~\bibnamefont {Sonoda}},\
  }\bibfield  {title} {\enquote {\bibinfo {title} {{Realization of Symmetry in
  the ERG Approach to Quantum Field Theory}},}\ }\href {\doibase
  10.1143/PTPS.181.1} {\bibfield  {journal} {\bibinfo  {journal}
  {Prog.~Theor.~Phys.~Suppl.}\ }\textbf {\bibinfo {volume} {181}},\ \bibinfo
  {pages} {1--166} (\bibinfo {year} {2010})},\ \Eprint
  {http://arxiv.org/abs/0909.0327} {arXiv:0909.0327 [hep-th]} \BibitemShut
  {NoStop}%
\bibitem [{\citenamefont {Pagani}(2016)}]{Pagani:2016pad}%
  \BibitemOpen
  \bibfield  {author} {\bibinfo {author} {\bibfnamefont {C.}~\bibnamefont
  {Pagani}},\ }\bibfield  {title} {\enquote {\bibinfo {title} {{Note on scaling
  arguments in the effective average action formalism}},}\ }\href {\doibase
  10.1103/PhysRevD.94.045001} {\bibfield  {journal} {\bibinfo  {journal} {Phys.
  Rev.}\ }\textbf {\bibinfo {volume} {D94}},\ \bibinfo {pages} {045001}
  (\bibinfo {year} {2016})},\ \Eprint {http://arxiv.org/abs/1603.07250}
  {arXiv:1603.07250 [hep-th]} \BibitemShut {NoStop}%
\bibitem [{\citenamefont {Wilson}(1969)}]{Wilson:1969zs}%
  \BibitemOpen
  \bibfield  {author} {\bibinfo {author} {\bibfnamefont {K.~G.}\ \bibnamefont
  {Wilson}},\ }\bibfield  {title} {\enquote {\bibinfo {title} {{Nonlagrangian
  models of current algebra}},}\ }\href {\doibase 10.1103/PhysRev.179.1499}
  {\bibfield  {journal} {\bibinfo  {journal} {Phys. Rev.}\ }\textbf {\bibinfo
  {volume} {179}},\ \bibinfo {pages} {1499--1512} (\bibinfo {year}
  {1969})}\BibitemShut {NoStop}%
\bibitem [{\citenamefont {Hughes}(1989)}]{Hughes:1988cp}%
  \BibitemOpen
  \bibfield  {author} {\bibinfo {author} {\bibfnamefont {J.}~\bibnamefont
  {Hughes}},\ }\bibfield  {title} {\enquote {\bibinfo {title} {{The OPE and the
  Exact Renormalization Group}},}\ }\href {\doibase
  10.1016/0550-3213(89)90025-4} {\bibfield  {journal} {\bibinfo  {journal}
  {Nucl. Phys.}\ }\textbf {\bibinfo {volume} {B312}},\ \bibinfo {pages}
  {125--154} (\bibinfo {year} {1989})}\BibitemShut {NoStop}%
\bibitem [{\citenamefont {Keller}\ and\ \citenamefont
  {Kopper}(1992)}]{Keller:1991bz}%
  \BibitemOpen
  \bibfield  {author} {\bibinfo {author} {\bibfnamefont {G.}~\bibnamefont
  {Keller}}\ and\ \bibinfo {author} {\bibfnamefont {C.}~\bibnamefont
  {Kopper}},\ }\bibfield  {title} {\enquote {\bibinfo {title} {{Perturbative
  renormalization of composite operators via flow equations. 1.}}}\ }\href
  {\doibase 10.1007/BF02096544} {\bibfield  {journal} {\bibinfo  {journal}
  {Commun. Math. Phys.}\ }\textbf {\bibinfo {volume} {148}},\ \bibinfo {pages}
  {445--468} (\bibinfo {year} {1992})}\BibitemShut {NoStop}%
\bibitem [{\citenamefont {Keller}\ and\ \citenamefont
  {Kopper}(1993)}]{Keller:1992by}%
  \BibitemOpen
  \bibfield  {author} {\bibinfo {author} {\bibfnamefont {G.}~\bibnamefont
  {Keller}}\ and\ \bibinfo {author} {\bibfnamefont {C.}~\bibnamefont
  {Kopper}},\ }\bibfield  {title} {\enquote {\bibinfo {title} {{Perturbative
  renormalization of composite operators via flow equations. 2. Short distance
  expansion}},}\ }\href {\doibase 10.1007/BF02096643} {\bibfield  {journal}
  {\bibinfo  {journal} {Commun. Math. Phys.}\ }\textbf {\bibinfo {volume}
  {153}},\ \bibinfo {pages} {245--276} (\bibinfo {year} {1993})}\BibitemShut
  {NoStop}%
\bibitem [{\citenamefont {Hollands}\ and\ \citenamefont
  {Kopper}(2012)}]{Hollands:2011gf}%
  \BibitemOpen
  \bibfield  {author} {\bibinfo {author} {\bibfnamefont {S.}~\bibnamefont
  {Hollands}}\ and\ \bibinfo {author} {\bibfnamefont {C.}~\bibnamefont
  {Kopper}},\ }\bibfield  {title} {\enquote {\bibinfo {title} {{The operator
  product expansion converges in perturbative field theory}},}\ }\href
  {\doibase 10.1007/s00220-012-1457-4} {\bibfield  {journal} {\bibinfo
  {journal} {Commun. Math. Phys.}\ }\textbf {\bibinfo {volume} {313}},\
  \bibinfo {pages} {257--290} (\bibinfo {year} {2012})},\ \Eprint
  {http://arxiv.org/abs/1105.3375} {arXiv:1105.3375 [hep-th]} \BibitemShut
  {NoStop}%
\bibitem [{\citenamefont {Holland}\ and\ \citenamefont
  {Hollands}(2015)}]{Holland:2014ifa}%
  \BibitemOpen
  \bibfield  {author} {\bibinfo {author} {\bibfnamefont {J.}~\bibnamefont
  {Holland}}\ and\ \bibinfo {author} {\bibfnamefont {S.}~\bibnamefont
  {Hollands}},\ }\bibfield  {title} {\enquote {\bibinfo {title} {{Recursive
  construction of operator product expansion coefficients}},}\ }\href {\doibase
  10.1007/s00220-014-2274-8} {\bibfield  {journal} {\bibinfo  {journal}
  {Commun. Math. Phys.}\ }\textbf {\bibinfo {volume} {336}},\ \bibinfo {pages}
  {1555--1606} (\bibinfo {year} {2015})},\ \Eprint
  {http://arxiv.org/abs/1401.3144} {arXiv:1401.3144 [math-ph]} \BibitemShut
  {NoStop}%
\bibitem [{\citenamefont {Holland}\ \emph {et~al.}(2016)\citenamefont
  {Holland}, \citenamefont {Hollands},\ and\ \citenamefont
  {Kopper}}]{Holland:2014pna}%
  \BibitemOpen
  \bibfield  {author} {\bibinfo {author} {\bibfnamefont {J.}~\bibnamefont
  {Holland}}, \bibinfo {author} {\bibfnamefont {S.}~\bibnamefont {Hollands}}, \
  and\ \bibinfo {author} {\bibfnamefont {C.}~\bibnamefont {Kopper}},\
  }\bibfield  {title} {\enquote {\bibinfo {title} {{The operator product
  expansion converges in massless $\varphi_{4}^{4}$- theory}},}\ }\href
  {\doibase 10.1007/s00220-015-2486-6} {\bibfield  {journal} {\bibinfo
  {journal} {Commun. Math. Phys.}\ }\textbf {\bibinfo {volume} {342}},\
  \bibinfo {pages} {385--440} (\bibinfo {year} {2016})},\ \Eprint
  {http://arxiv.org/abs/1411.1785} {arXiv:1411.1785 [hep-th]} \BibitemShut
  {NoStop}%
\bibitem [{\citenamefont {Fr{\"{o}}b}\ \emph {et~al.}(2016)\citenamefont
  {Fr{\"{o}}b}, \citenamefont {Holland},\ and\ \citenamefont
  {Hollands}}]{Frob:2015uqy}%
  \BibitemOpen
  \bibfield  {author} {\bibinfo {author} {\bibfnamefont {M.~B.}\ \bibnamefont
  {Fr{\"{o}}b}}, \bibinfo {author} {\bibfnamefont {J.}~\bibnamefont {Holland}},
  \ and\ \bibinfo {author} {\bibfnamefont {S.}~\bibnamefont {Hollands}},\
  }\bibfield  {title} {\enquote {\bibinfo {title} {{All-order bounds for
  correlation functions of gauge-invariant operators in Yang-Mills theory}},}\
  }\href {\doibase 10.1063/1.4967747} {\bibfield  {journal} {\bibinfo
  {journal} {J. Math. Phys.}\ }\textbf {\bibinfo {volume} {57}},\ \bibinfo
  {pages} {122301} (\bibinfo {year} {2016})},\ \Eprint
  {http://arxiv.org/abs/1511.09425} {arXiv:1511.09425 [math-ph]} \BibitemShut
  {NoStop}%
\bibitem [{\citenamefont {Fr{\"{o}}b}\ and\ \citenamefont
  {Holland}(2016)}]{Frob:2016mzv}%
  \BibitemOpen
  \bibfield  {author} {\bibinfo {author} {\bibfnamefont {M.~B.}\ \bibnamefont
  {Fr{\"{o}}b}}\ and\ \bibinfo {author} {\bibfnamefont {J.}~\bibnamefont
  {Holland}},\ }\bibfield  {title} {\enquote {\bibinfo {title} {{All-order
  existence of and recursion relations for the operator product expansion in
  Yang-Mills theory}},}\ }\href@noop {} {\  (\bibinfo {year} {2016})},\ \Eprint
  {http://arxiv.org/abs/1603.08012} {arXiv:1603.08012 [math-ph]} \BibitemShut
  {NoStop}%
\bibitem [{\citenamefont {Zimmermann}(1973)}]{Zimmermann:1972tv}%
  \BibitemOpen
  \bibfield  {author} {\bibinfo {author} {\bibfnamefont {W.}~\bibnamefont
  {Zimmermann}},\ }\bibfield  {title} {\enquote {\bibinfo {title} {{Normal
  products and the short distance expansion in the perturbation theory of
  renormalizable interactions}},}\ }\bibfield  {booktitle} {\emph {\bibinfo
  {booktitle} {{In *Tegernsee 1998, Quantum field theory* 278-309}}},\ }\href
  {\doibase 10.1016/0003-4916(73)90430-2} {\bibfield  {journal} {\bibinfo
  {journal} {Annals Phys.}\ }\textbf {\bibinfo {volume} {77}},\ \bibinfo
  {pages} {570--601} (\bibinfo {year} {1973})},\ \bibinfo {note} {[Lect. Notes
  Phys.558,278(2000)]}\BibitemShut {NoStop}%
\bibitem [{\citenamefont {Wilson}\ and\ \citenamefont
  {Kogut}(1974)}]{Wilson:1973jj}%
  \BibitemOpen
  \bibfield  {author} {\bibinfo {author} {\bibfnamefont {K.~G.}\ \bibnamefont
  {Wilson}}\ and\ \bibinfo {author} {\bibfnamefont {J.~B.}\ \bibnamefont
  {Kogut}},\ }\bibfield  {title} {\enquote {\bibinfo {title} {{The
  renormalization group and the epsilon expansion}},}\ }\href@noop {}
  {\bibfield  {journal} {\bibinfo  {journal} {Phys.~Rept.}\ }\textbf {\bibinfo
  {volume} {12}},\ \bibinfo {pages} {75--200} (\bibinfo {year}
  {1974})}\BibitemShut {NoStop}%
\bibitem [{\citenamefont {Polchinski}(1984)}]{Polchinski:1983gv}%
  \BibitemOpen
  \bibfield  {author} {\bibinfo {author} {\bibfnamefont {J.}~\bibnamefont
  {Polchinski}},\ }\bibfield  {title} {\enquote {\bibinfo {title}
  {{Renormalization and Effective Lagrangians}},}\ }\href {\doibase
  10.1016/0550-3213(84)90287-6} {\bibfield  {journal} {\bibinfo  {journal}
  {Nucl. Phys.}\ }\textbf {\bibinfo {volume} {B231}},\ \bibinfo {pages}
  {269--295} (\bibinfo {year} {1984})}\BibitemShut {NoStop}%
\bibitem [{\citenamefont {Morris}(1994{\natexlab{a}})}]{Morris:1993qb}%
  \BibitemOpen
  \bibfield  {author} {\bibinfo {author} {\bibfnamefont {T.~R.}\ \bibnamefont
  {Morris}},\ }\bibfield  {title} {\enquote {\bibinfo {title} {{The exact
  renormalization group and approximate solutions}},}\ }\href {\doibase
  10.1142/S0217751X94000972} {\bibfield  {journal} {\bibinfo  {journal} {Int.
  J. Mod. Phys.}\ }\textbf {\bibinfo {volume} {A9}},\ \bibinfo {pages}
  {2411--2450} (\bibinfo {year} {1994}{\natexlab{a}})},\ \Eprint
  {http://arxiv.org/abs/hep-ph/9308265} {arXiv:hep-ph/9308265 [hep-ph]}
  \BibitemShut {NoStop}%
\bibitem [{\citenamefont {Wetterich}(1993)}]{Wetterich:1992yh}%
  \BibitemOpen
  \bibfield  {author} {\bibinfo {author} {\bibfnamefont {C.}~\bibnamefont
  {Wetterich}},\ }\bibfield  {title} {\enquote {\bibinfo {title} {{Exact
  evolution equation for the effective potential}},}\ }\href {\doibase
  10.1016/0370-2693(93)90726-X} {\bibfield  {journal} {\bibinfo  {journal}
  {Phys. Lett.}\ }\textbf {\bibinfo {volume} {B301}},\ \bibinfo {pages}
  {90--94} (\bibinfo {year} {1993})}\BibitemShut {NoStop}%
\bibitem [{\citenamefont {Ellwanger}(1994)}]{Ellwanger:1993mw}%
  \BibitemOpen
  \bibfield  {author} {\bibinfo {author} {\bibfnamefont {U.}~\bibnamefont
  {Ellwanger}},\ }\bibfield  {title} {\enquote {\bibinfo {title} {{FLow
  equations for N point functions and bound states}},}\ }\bibfield  {booktitle}
  {\emph {\bibinfo {booktitle} {{Proceedings, Workshop on Quantum field
  theoretical aspects of high energy physics: Bad Frankenhausen, Germany,
  September 20-24, 1993}}},\ }\href {\doibase 10.1007/BF01555911} {\bibfield
  {journal} {\bibinfo  {journal} {Z. Phys.}\ }\textbf {\bibinfo {volume}
  {C62}},\ \bibinfo {pages} {503--510} (\bibinfo {year} {1994})},\ \bibinfo
  {note} {[,206(1993)]},\ \Eprint {http://arxiv.org/abs/hep-ph/9308260}
  {arXiv:hep-ph/9308260 [hep-ph]} \BibitemShut {NoStop}%
\bibitem [{\citenamefont {Igarashi}\ \emph {et~al.}(2016)\citenamefont
  {Igarashi}, \citenamefont {Itoh},\ and\ \citenamefont
  {Sonoda}}]{Igarashi:2016qdr}%
  \BibitemOpen
  \bibfield  {author} {\bibinfo {author} {\bibfnamefont {Y.}~\bibnamefont
  {Igarashi}}, \bibinfo {author} {\bibfnamefont {K.}~\bibnamefont {Itoh}}, \
  and\ \bibinfo {author} {\bibfnamefont {H.}~\bibnamefont {Sonoda}},\
  }\bibfield  {title} {\enquote {\bibinfo {title} {{On the wave function
  renormalization for Wilson actions and their one particle irreducible
  actions}},}\ }\href {\doibase 10.1093/ptep/ptw121} {\bibfield  {journal}
  {\bibinfo  {journal} {PTEP}\ }\textbf {\bibinfo {volume} {2016}},\ \bibinfo
  {pages} {093B04} (\bibinfo {year} {2016})},\ \Eprint
  {http://arxiv.org/abs/1607.01521} {arXiv:1607.01521 [hep-th]} \BibitemShut
  {NoStop}%
\bibitem [{\citenamefont {Sonoda}(2015)}]{Sonoda:2015bla}%
  \BibitemOpen
  \bibfield  {author} {\bibinfo {author} {\bibfnamefont {H.}~\bibnamefont
  {Sonoda}},\ }\bibfield  {title} {\enquote {\bibinfo {title} {{Equivalence of
  Wilson Actions}},}\ }\href {\doibase 10.1093/ptep/ptv130} {\bibfield
  {journal} {\bibinfo  {journal} {PTEP}\ }\textbf {\bibinfo {volume} {2015}},\
  \bibinfo {pages} {103B01} (\bibinfo {year} {2015})},\ \Eprint
  {http://arxiv.org/abs/1503.08578} {arXiv:1503.08578 [hep-th]} \BibitemShut
  {NoStop}%
\bibitem [{\citenamefont {Morris}(1994{\natexlab{b}})}]{Morris:1994ie}%
  \BibitemOpen
  \bibfield  {author} {\bibinfo {author} {\bibfnamefont {T.~R.}\ \bibnamefont
  {Morris}},\ }\bibfield  {title} {\enquote {\bibinfo {title} {{Derivative
  expansion of the exact renormalization group}},}\ }\href {\doibase
  10.1016/0370-2693(94)90767-6} {\bibfield  {journal} {\bibinfo  {journal}
  {Phys. Lett.}\ }\textbf {\bibinfo {volume} {B329}},\ \bibinfo {pages}
  {241--248} (\bibinfo {year} {1994}{\natexlab{b}})},\ \Eprint
  {http://arxiv.org/abs/hep-ph/9403340} {arXiv:hep-ph/9403340 [hep-ph]}
  \BibitemShut {NoStop}%
\bibitem [{\citenamefont {Berges}\ \emph {et~al.}(2002)\citenamefont {Berges},
  \citenamefont {Tetradis},\ and\ \citenamefont {Wetterich}}]{Berges:2000ew}%
  \BibitemOpen
  \bibfield  {author} {\bibinfo {author} {\bibfnamefont {J.}~\bibnamefont
  {Berges}}, \bibinfo {author} {\bibfnamefont {N.}~\bibnamefont {Tetradis}}, \
  and\ \bibinfo {author} {\bibfnamefont {C.}~\bibnamefont {Wetterich}},\
  }\bibfield  {title} {\enquote {\bibinfo {title} {{Nonperturbative
  renormalization flow in quantum field theory and statistical physics}},}\
  }\href {\doibase 10.1016/S0370-1573(01)00098-9} {\bibfield  {journal}
  {\bibinfo  {journal} {Phys. Rept.}\ }\textbf {\bibinfo {volume} {363}},\
  \bibinfo {pages} {223--386} (\bibinfo {year} {2002})},\ \Eprint
  {http://arxiv.org/abs/hep-ph/0005122} {arXiv:hep-ph/0005122 [hep-ph]}
  \BibitemShut {NoStop}%
\bibitem [{\citenamefont {Sonoda}(2017)}]{Sonoda:2017rro}%
  \BibitemOpen
  \bibfield  {author} {\bibinfo {author} {\bibfnamefont {H.}~\bibnamefont
  {Sonoda}},\ }\bibfield  {title} {\enquote {\bibinfo {title} {{The generating
  functional of correlation functions as a high momentum limit of a Wilson
  action}},}\ }\href@noop {} {\  (\bibinfo {year} {2017})},\ \Eprint
  {http://arxiv.org/abs/1706.00198} {arXiv:1706.00198 [hep-th]} \BibitemShut
  {NoStop}%
\bibitem [{\citenamefont {Rosten}(2012)}]{Rosten:2010vm}%
  \BibitemOpen
  \bibfield  {author} {\bibinfo {author} {\bibfnamefont {O.~J.}\ \bibnamefont
  {Rosten}},\ }\bibfield  {title} {\enquote {\bibinfo {title} {{Fundamentals of
  the Exact Renormalization Group}},}\ }\href {\doibase
  10.1016/j.physrep.2011.12.003} {\bibfield  {journal} {\bibinfo  {journal}
  {Phys. Rept.}\ }\textbf {\bibinfo {volume} {511}},\ \bibinfo {pages}
  {177--272} (\bibinfo {year} {2012})},\ \Eprint
  {http://arxiv.org/abs/1003.1366} {arXiv:1003.1366 [hep-th]} \BibitemShut
  {NoStop}%
\bibitem [{\citenamefont {Blaizot}\ \emph {et~al.}(2006)\citenamefont
  {Blaizot}, \citenamefont {Mendez-Galain},\ and\ \citenamefont
  {Wschebor}}]{Blaizot:2005wd}%
  \BibitemOpen
  \bibfield  {author} {\bibinfo {author} {\bibfnamefont {J.-P.}\ \bibnamefont
  {Blaizot}}, \bibinfo {author} {\bibfnamefont {R.}~\bibnamefont
  {Mendez-Galain}}, \ and\ \bibinfo {author} {\bibfnamefont {N.}~\bibnamefont
  {Wschebor}},\ }\bibfield  {title} {\enquote {\bibinfo {title} {{Non
  perturbative renormalisation group and momentum dependence of n-point
  functions (I)}},}\ }\href {\doibase 10.1103/PhysRevE.74.051116} {\bibfield
  {journal} {\bibinfo  {journal} {Phys. Rev.}\ }\textbf {\bibinfo {volume}
  {E74}},\ \bibinfo {pages} {051116} (\bibinfo {year} {2006})},\ \Eprint
  {http://arxiv.org/abs/hep-th/0512317} {arXiv:hep-th/0512317 [hep-th]}
  \BibitemShut {NoStop}%
\bibitem [{\citenamefont {Osborn}(2015)}]{Osborn:2015}%
  \BibitemOpen
  \bibfield  {author} {\bibinfo {author} {\bibfnamefont {H.}~\bibnamefont
  {Osborn}},\ }\href@noop {} {\enquote {\bibinfo {title} {Functional
  representations of conformal symmetry in quantum field theory},}\ } (\bibinfo
  {year} {2015}),\ \bibinfo {note} {unpublished}\BibitemShut {NoStop}%
\bibitem [{\citenamefont {Ball}\ \emph {et~al.}(1995)\citenamefont {Ball},
  \citenamefont {Haagensen}, \citenamefont {Latorre},\ and\ \citenamefont
  {Moreno}}]{Ball:1994ji}%
  \BibitemOpen
  \bibfield  {author} {\bibinfo {author} {\bibfnamefont {R.~D.}\ \bibnamefont
  {Ball}}, \bibinfo {author} {\bibfnamefont {P.~E.}\ \bibnamefont {Haagensen}},
  \bibinfo {author} {\bibfnamefont {J.~I.}\ \bibnamefont {Latorre}}, \ and\
  \bibinfo {author} {\bibfnamefont {E.}~\bibnamefont {Moreno}},\ }\bibfield
  {title} {\enquote {\bibinfo {title} {{Scheme independence and the exact
  renormalization group}},}\ }\href {\doibase 10.1016/0370-2693(95)00025-G}
  {\bibfield  {journal} {\bibinfo  {journal} {Phys. Lett.}\ }\textbf {\bibinfo
  {volume} {B347}},\ \bibinfo {pages} {80--88} (\bibinfo {year} {1995})},\
  \Eprint {http://arxiv.org/abs/hep-th/9411122} {arXiv:hep-th/9411122 [hep-th]}
  \BibitemShut {NoStop}%
\end{thebibliography}%

\end{document}